\documentclass[12pt]{iopart}
\usepackage[utf8]{inputenc}
\usepackage[english]{babel}
\usepackage{graphicx}
\usepackage{afterpage}
\usepackage{lipsum}
\usepackage{pdflscape}
\usepackage{url}
\usepackage{paralist}
\graphicspath{ {images/} }
\usepackage{caption}
\usepackage{iopams}  
\usepackage[breaklinks=true,colorlinks=true,linkcolor=blue,urlcolor=blue,citecolor=blue]{hyperref}
\usepackage{geometry}
\usepackage{tabularx}
\usepackage{xfrac}

\begin{document}

\title{Complex Eigenvalues in a pseudo-Hermitian $\beta$-Laguerre ensemble}

\author{Cleverson Andrade Goulart$^1$, Gleb Oshanin$^2$,  and Mauricio Porto Pato$^1$}

\address{$^1$Instituto de Física, Universidade de São Paulo, Caixa Postal 66318, São Paulo 05314-970, SP, Brazil}
\address{$^2$Laboratoire de Physique Th\'eorique de la Mati\`{e}re Condens\'ee (UMR	CNRS 7600), Sorbonne Universit\'e, 4 place Jussieu, Paris, 75005, France.}
%\address{$^3$B. Verkin Institute for Low Temperature Physics and Engineering, 47 Nauky Ave., Kharkiv 61103,Ukraine}
%\address{$^4$ King's College London, Strand, London WC2R 2LS, UK}

%\ead{}
\vspace{10pt}
%\begin{indented}
%\item[]August 2024 
%\end{indented}

\begin{abstract}

 Non-Hermitian PT-symmetric models have been extensively studied in recent years. Following the seminal work that reduced classical random matrix ensembles to a tridiagonal form, several efforts have aimed to generalize this framework to non-Hermitian extensions of the so-called $\beta$-ensembles. In particular, while the transition of eigenvalues from the real axis to the complex plane has been well characterized for the $\beta$-Hermite ensemble under symmetry breaking, the behavior of the $\beta$-Laguerre ensemble in a similar non-Hermitian setting remains less understood. In this work, we investigate an ensemble of unstable matrices isospectral to the $\beta$-Laguerre ensemble. Introducing a small non-Hermitian perturbation breaks the symmetry and drives the eigenvalues into the complex plane. We derive analytical expressions for the loci of complex-conjugate eigenvalue pairs, which organize into a balloon-like structure in the complex plane, followed by a discrete finite line of real eigenvalues. The asymptotic behavior of these eigenvalues is analyzed in the large matrix-size limit, and our theoretical predictions are supported by numerical simulations.
\end{abstract}

\vspace{2pc}
\noindent{\it Keywords}: random matrix theory, $\beta$-Laguerre ensemble, pseudo-Hermitian, PT-symmetry, self-averaging system.
%\maketitle

\section{Introduction}

\setlength{\mathindent}{50pt}

Random matrices first appeared in statistics in Wishart’s work on multivariate analysis, where he introduced sample covariance matrices. Later, in the 1950s, Wigner employed random matrices in physics to model the statistical properties of energy levels in heavy nuclei. These seminal works gave rise to the study of Wishart (or Laguerre) ensembles and Wigner ensembles, which remain central objects in random matrix theory.
A key feature of these ensembles is their classification according to underlying symmetry. Dyson introduced the notion of the Dyson index $\beta$, which takes the values $\beta = 1, 2,$ or $4$ depending on whether the matrix entries are real, complex, or quaternionic. In the case of Gaussian ensembles, this leads to the classical Gaussian Orthogonal Ensemble (GOE, $\beta=1$), Gaussian Unitary Ensemble (GUE, $\beta=2$), and Gaussian Symplectic Ensemble (GSE, $\beta=4$).The introduction of non-Hermitian matrices has revealed qualitatively new phenomena. The Ginibre ensemble \cite{ginibre1965statistical}, consisting of matrices with independent Gaussian entries, exhibits eigenvalues distributed across the complex plane according to the circular law in the large-$N$ limit \cite{shivam2023many, byun2025progress}. In parallel, Dumitriu and Edelman, in their seminal paper \cite{dumitriu2002matrix}, developed a tridiagonal representation of classical Gaussian and Wishart ensembles, allowing the Dyson index $\beta$ to take arbitrary continuous values and giving rise to $\beta$-ensembles. These $\beta$-ensembles preserve the key spectral laws of their full-matrix counterparts in the Hermitian case. 
Extensions of RMT to non-Hermitian and pseudo-Hermitian matrices have attracted increasing interest. In particular, we can mention the pioneering work of Pato and Bohigas, who studied the conditions under which eigenvalues remain real after relaxing the Hermitian constraint \cite{bohigas2013non}, thereby using pseudo-Hermitian matrices. An $\eta$-pseudo-Hermitian matrix $A$ satisfies \cite{mostafazadeh2002pseudo}
\begin{equation}
	\label{pseudo}
	A=\eta^{-1}A^{\dagger}\eta,
\end{equation}
where $\eta = \eta^{\dagger}$ is a Hermitian invertible matrix. In \cite{mostafazadeh2002pseudo} it was shown that under the pseudo-Hermiticity condition given in eq. (\ref{pseudo}), the eigenvalues are either real or occur in complex-conjugate pairs. Furthermore, it was demonstrated that the reality of the spectrum can be established through the introduction of an auxiliary matrix,
\begin{eqnarray}
	\eta^{1/2}\cdot \eta^{1/2} &=& \eta,\\
	(\eta^{1/2})^{-1} &=& \eta^{-1/2},
\end{eqnarray}
provided that both $\eta^{1/2}$ and its inverse exist. This representation entails a Hermitian matrix $H$ of the form
\begin{equation}
	H = \eta^{1/2} A \eta^{-1/2} = \eta^{-1/2} [\eta A \eta^{-1}] \eta^{1/2} = \eta^{-1/2} A^{\dagger} \eta^{1/2} = H^{\dagger} ,
\end{equation}	
implying that $A$ and $H$ share the same set of real eigenvalues. Thus, the reality of the spectrum follows directly from the Hermitian nature of $H$.

Such pseudo-Hermitian structures have been applied in physics, including population biology, active metamaterials, and non-Hermitian quantum systems \cite{nelson1998non, ghatak2020observation, shamshutdinova2008feshbach, zhu2014pt}. In particular, when the pseudo-Hermitian condition is imposed on $\beta$-ensembles, the eigenvalues follow Wigner’s semicircle law \cite{forrester2010log} in the Gaussian case and the Marchenko–Pastur distribution \cite{MP} in the Wishart case in the large-$N$ limit. The introduction of non-Hermiticity, however, is accompanied by a doubling of Dyson’s $\beta$ index when the behavior of the trace is considered in the joint distribution of matrix elements \cite{goulart2023double}.
The study of the non-Hermitian $\beta$-Hermite ensemble with real eigenvalues is part of a broader series of works introduced through the physical model of non-Hermitian random Schrodinger operators. This model, developed by Hatano and Nelson \cite{hatano1996localization, hatano1997vortex}, serves as a framework for studying systems subject to a constant imaginary vector potential. In their investigation of magnetic flux lines, the authors uncovered an intriguing phenomenon related to the eigenvector localization. Specifically, they argued that when the eigenvectors are localized, the eigenvalues remain real, whereas in the delocalized case, the eigenvalues become complex. This pioneering work sparked an active field of research. 

Rank-one perturbations provide a natural and analytically tractable framework for exploring spectral deformations. For example, such deformations of the Gaussian ensemble serve as an analytic tool for studying the statistics of resonances in quantum scattering from chaotic domains. The density of eigenvalues in these scenarios is a central topic of interest in both physics and mathematics \cite{fyodorov2022extreme,fyodorov2003random}. In the Hermitian setting, they induce interlacing eigenvalue structures \cite{tao2012topics}. In the non-Hermitian setting, however, the spectral behavior is more subtle and less understood, particularly for ensembles where the unperturbed matrix has real, positive eigenvalues. A class of non-Hermitian tridiagonal matrices with complex eigenvalues has been explored in a series of works by Goldsheid and Khoruzhenko \cite{goldsheid1998distribution, goldsheid2000eigenvalue}. Although their research was not initially motivated by a connection to the $\beta$-ensembles, they describe the eigenvalue curves associated with non-Hermitian operators with boundary conditions. Their analysis relies on the Lyapunov exponent and the integrated density of states of a related symmetric eigenvalue problem. Similarly, in the context of $\beta$-ensemble investigations, Kozhan used an imaginary potential to model an effective random Hamiltonian and computed the joint eigenvalue density of a rank-one non-Hermitian perturbation of the Gaussian and Laguerre $\beta$-ensembles \cite{kozhan2017rank}. Kozhan's work demonstrated that the deformations introduced by such a perturbation led to integrable random ensembles, meaning that the eigenvalue correlations could be computed in closed form. These models reveal that such perturbations can produce significant changes in the eigenvalue distribution, such as the emergence of complex-conjugate eigenvalue pairs and anisotropic spreading in the complex plane.
In light of the rank-one perturbation in the context of the $\beta$-Hermite ensemble \cite{marinello2016pseudo}, a natural question  concerns the non-Hermitian matrices that are isospectral (i.e., have the same set of eigenvalues) to other $\beta$-ensembles. Some progress has been made in this direction with a recent publication on non-Hermitian random matrices \cite{patopseudo}, where some general features of non-Hermitian $\beta$-ensembles have been discussed. However, a detailed analysis of the properties of spectra for the $\beta$-Laguerre ensemble is still missing at present.

Motivated by the analysis of spectral properties of the $\beta$-Hermite ensemble, 
in the present paper we study analogous properties of a set of unstable matrices that are isospectral to the $\beta$-Laguerre ensemble and discuss the behavior of its eigenvalues in the complex plane. Calculating first the roots the \textit{mean} characteristic polynomial, we determine the loci of pairs of complex-conjugate eigenvalues. We show that these pairs form on the complex plane a kind of a bounded anisotropic "balloon"  which closes at a certain threshold value, beyond which the imaginary part of eigenvalues vanishes such that the balloon structure is followed by a discrete set of real eigenvalues. We present an asymptotic analysis of geometric characteristics of the observed baloon structure.
Next, we access the validity of such an analysis by determining  the variance of the characteristic polynomial.  On this basis, we show that the system is self-averaging  \cite{pastur} and fluctuations around the mean values of the eigenvalues become progressively less important when the matrix dimension tends to infinity. 
This signifies that the above description is completely justified, and also permits us to quantify anisotropic fluctuations of eigenvalues beyond their mean values. Our numerical simulations completely validate the presented picture.
The paper is organized as follows. In Sec. 2, we introduce the unstable pseudo-Hermitian model for the $\beta$-Laguerre ensemble and discuss the reality of the spectrum under a particular choice of the diagonal Hermitian operator $\eta$ that satisfies eq. (\ref{pseudo}). In Sec. 2.2, we consider the complex eigenvalues resulting from a non-Hermitian perturbation and highlight some properties observed in the spectrum as functions of the model parameters.
Since the traditional framework of averaging the resolvent proves to be unsuitable in this context, Sec. 3 presents an alternative analytical approach based on the eigenvalues of the mean characteristic polynomial. This method leverages the asymptotic form of Laguerre polynomials in the complex domain as a key technique. To validate this approach—motivated by the convergence to a deterministic function in the Hermitian case—we include an analysis of the model’s relative variance in the asymptotic regime, i.e., as $N \rightarrow \infty$, in Sec. 4. The main conclusions are presented in Sec. 5, along with some perspectives for extending the present work.

\section{The pseudo-Hermitian model} 

In this Section we present our model. In order to make this Section self-contained, we first
	recall the definition of the standard Hermitian $\beta$-Laguerre ensembles and briefly discuss their properties. 
	Then, we show how the Hermitian condition can be relaxed, giving rise to pseudo-Hermitian $\beta$-Laguerre ensembles with real eigenvalues. Lastly, we introduce a perturbation of the latter ensemble by adding corner elements, which define the model of a pseudo-Hermitian $\beta$-Laguerre ensemble (see eq. (\ref{main})). 
	Since the properties of complex eigenvalues of this ensemble are of focal interest in our paper, we present some numerical calculations to illustrate the properties of the constructed pseudo-Hermitian model and use them to gain some intuition about the model.

\subsection[Real Eigenvalues]{Pseudo-Hermitian $\beta$-Laguerre ensemble with real eigenvalues }

We begin with a brief reminder on the so-called $\beta$-Laguerre ensemble \cite{dumitriu2002matrix}. 
Consider a rectangular ($N \times M$) bi-diagonal matrix $B$ with elements $(a_{N},a_{N-1},\dots, a_{1})$ on the diagonal  
and elements $( b_{N-1}, b_{N-2},\dots, b_{1})$ on the lower sub-diagonal, which are all  sorted independently from the distribution, 
\begin{equation}
	\label{disF}
	f_{\nu}(x)=\frac{2^{1-\frac{\nu}{2}}}{\Gamma(\nu/2)}x^{\nu-1}\exp\left[-\frac{x^{2}}{2}\right], \quad x\geq 0
\end{equation}
where $\nu=\beta(M-i+1)$ for the diagonal elements and $\nu=\beta(N-j)$ for the sub-diagonal ones. 
The joint distribution of the elements is given by
\begin{equation}
	\label{PB}
	P(B_{\beta})=\frac{1}{Z}\prod\limits_{i=1}^{N}a_{N-i+1}^{\nu-1}\exp\left(-\frac{1}{2}a_{N-i+1}^{2}\right)\prod\limits_{j=1}^{N-1}b_{N-j}^{\nu-1}\exp\left(-\frac{1}{2}b_{N-j}^{2}\right),
\end{equation}
where $Z$ is the normalization constant. The $\beta$-Laguerre ensemble is then 
defined from the matrix $B$ as the square tridiagonal matrix ($N \times N$), 	
\[
\label{3}
L_{\beta}=B.B^{T}=
\left[
\begin{array}{cccc}
	a_{N}^{2} & a_{N}b_{N-1} &  &  \\
	a_{N}b_{N-1} & a_{N-1}^{2}+b_{N-1}^{2} & \ddots &  \\
	& \ddots & \ddots & a_{2}b_{1} \\
	&  & a_{2}b_{1} & a_{1}^{2}+b_{1}^{2}
\end{array}
\right].
\]
The Jacobian $J_{B\rightarrow L}$ of the transformation from the bidiagonal matrix B to the tridiagonal matrix $L_{\beta}$ has a  well-known form \cite{dumitriu2002matrix}, 
\begin{equation}
	J_{B\rightarrow L}=\left[2^{N}a_{1}\prod\limits_{i=2}^{N}a_{i}^{2}\right]^{-1}.
\end{equation}
permitting us to cast the distribution of the elements in  eq. (\ref{PB}) into the one 
which applies to the elements of the matrix $L_{\beta}$ :
\begin{eqnarray*}
	P(L_{\beta})&=\frac{1}{Z'}2^{-N}a_{1}^{\beta(M-N+1)-2}\exp\left(-\frac{1}{2}\Tr L_{\beta}\right)\\ &\times \prod\limits_{i=1}^{N-1}a_{i+1}^{\beta(M-N+1+i)-3}
	\times \prod\limits_{j=1}^{N-1}b_{j}^{j\beta-1}.
\end{eqnarray*}
Finally, using the above distribution, one constructs the joint distribution of the eigenvalues of the $L_{\beta}$ ensemble \cite{dumitriu2002matrix} :
\begin{eqnarray*}
		P(\lambda_{1},\lambda_{2},\dots,\lambda_{N})&=Z_{NM}\exp\left(-\frac{1}{2}\sum\limits_{i=1}^{N}\lambda_{i}\right)\\
		&\times \prod\limits_{i=1}^{N}\lambda_{i}^{\frac{\beta}{2}(M-N+1)-1}\prod\limits_{i\neq j}|\lambda_{i}-\lambda_{j}|^{\beta} .
\end{eqnarray*}
We note parenthetically that in the limit $N, M \rightarrow \infty$ with the ratio $c=N/M$ kept fixed, the above density of the eigenvalues converges to  the Marchenko-Pastur distribution \cite{MP} (see also \cite{forrester2010log}),
\begin{equation}
	\rho(\lambda)=\frac{1}{2\pi\beta\lambda}\sqrt{(\lambda_{+}-\lambda)(\lambda-\lambda_{-})},
\end{equation}
where $\lambda_{\pm}= N\beta(c^{-1/2}\pm 1)^{2}$. 
%When the size of the matrix becomes large� approaching the asymptotic limit—as $N$ and $M$ increase, the quotient %remains fixed at a constant value, $c$.

Next, we want to relax the Hermitian condition, which can be done in a number of different ways. Here we do it  by somewhat modifying the matrix B. More specifically, we construct two new matrices $B_{\pm}$ 
 such that the diagonal elements of $B_{\pm}$ remain unchanged and are identical to those of $B$, while the off-diagonal elements are given by $( b_{N-1}^{1\pm \alpha}, b_{N-2}^{1\pm \alpha},\dots, b_{1}^{1\pm \alpha})$, where 
 $\alpha$ is a real number.  The particular choice of this parameter will be addressed in greater detail in the subsequent section. 
 Then, we define a new tridiagonal matrix $L(\alpha)$ as
\[
L(\alpha)=B_{-}B_{+}^{T}=\left[\begin{array}{cccc}
	a_{N}^{2} & a_{N}b_{N-1}^{1+\alpha} & \dots & 0 \\
	a_{N}b_{N-1}^{1-\alpha}	
	& a_{N-1}^{2}+b_{N-1}^{2} & \ddots & \vdots \\
	\vdots & \ddots & \ddots & a_{2}b_{1}^{1+\alpha}	
	\\
	0 & \dots & a_{2}b_{1}^{1-\alpha}	
	
	& a_{1}^{2}+b_{1}^{2}
\end{array}\right].
\] 
This new matrix is pseudo-Hermitian, which can easily be demonstrated 
by defining a diagonal matrix $\eta$ with elements,
\begin{equation}
	\label{eta}
	\mbox{diag}(\eta)=\left(1,b_{N-1}^{2\alpha},\dots,\prod\limits_{k=1}^{N-1}(b_{N-k})^{2\alpha}\right),
\end{equation}
so that equation $L=\eta^{-1}L^{\dagger}\eta$ is automatically satisfied.\\
Since the elements of $\eta$ are diagonal and positive, its square root can be constructed as
\begin{equation}
	\mbox{diag}(\eta^{1/2})=\left(1,\sqrt{b_{N-1}^{2\alpha}},\dots,\prod\limits_{k=1}^{N-1}\sqrt{(b_{N-k})^{2\alpha}}\right),
\end{equation} 
such that the similarity transformation $H=\eta(\alpha)^{1/2}L(\alpha)\eta(\alpha)^{-1/2}$ of $L(\alpha)$ produces a Hermitian matrix,
\[
H=\left[\begin{array}{cccc}
	a_{N}^{2} & a_{N}b_{N-1} & \dots & 0 \\
	a_{N}b_{N-1} & a_{N-1}^{2}+b_{N-1}^{2}
	& \ddots & \vdots \\
	\vdots & \ddots & \ddots & a_{2}b_{1} \\
	0 & \dots & a_{2}b_{1}
	& a_{1}^{2}+b_{1}^{2}
\end{array}\right],
\]
that belongs to the $\beta$-Laguerre ensemble. Thus, it turns out that $L(\alpha)$ becomes a family of matrices which are pseudo-Hermitian with the same real eigenvalues of the $\beta$-Laguerre ensemble. In physical terms, the above arguments indicate that a pseudo-Hermiticity does not necessarily lead to a specific spectral statistics. The similarity transformation given by eq. (\ref{pseudo}) can be regarded as a generalized gauge transformation.
	
	\subsection[Complex eigenvalues]{Pseudo-Hermitian model with complex eigenvalues}
	
The reality of the spectrum for a Hermitian operator is a well-known result found in linear algebra. However, Hermiticity is not a necessary condition. A quasi-Hermitian operator $A$, defined by the relation $\xi A = A^{\dagger}\xi$, where $\xi$ is any positive-definite and bounded operator, possesses a real spectrum \cite{williams1969operators}. It can be observed that this requirement is stronger than that imposed by eq. (\ref{pseudo}), which holds for any invertible operator $\eta$.

From the distribution given in eq. (\ref{disF}), it is possible to compute the $k-th$ moment of the random variable $b_{N-j}$,
\begin{equation}
	\langle b_{N-j}^{k}\rangle=\frac{2^{1-\frac{\beta(N-j)}{2}}}{\Gamma(\frac{\beta(N-j)}{2})}\int\limits_{0}^{\infty} b^{\beta(N-j) +1+k}\exp(-\frac{b^{2}}{2}) db=2^{k/2}\frac{\Gamma (\frac{\beta(N-j) + k}{2})}{\Gamma (\frac{\beta(N-j)}{2})}
\end{equation} 	
Since each element of $B_{\pm}$ is sorted independently from the distribution in eq. (\ref{disF}),
\begin{equation}
		\langle\eta_{NN}\rangle =\langle\prod\limits_{j=1}^{N-1}(b_{N-j})^{2\alpha}\rangle=\prod\limits_{j=1}^{N-1} \langle (b_{N-j})^{2\alpha} \rangle = \prod\limits_{j=1}^{N-1} 2^{\alpha}\frac{\Gamma (\frac{\beta (N-j)}{2} + \alpha)}{\Gamma(\frac{\beta (N-j)}{2})},
\end{equation}
which suggests that as the matrix dimensions increase, the norm of the average of $\eta$ also increases. Due to the diagonal structure of $\eta$, this operator becomes unbounded in the asymptotic limit of large $N$, which in turn causes the matrix $L(\alpha)$ to become unstable. It is important to note that for two Hilbert spaces $\mathcal{H}_{1}$ and $\mathcal{H}_{2}$, each with its own inner product $\langle \cdot | \cdot \rangle_{1,2}$, and a linear operator $O$ that maps $\mathcal{H}_{1}$ to $\mathcal{H}_{2}$, the operator $O$ is bounded if it satisfies the inequality $\| O\Psi \|_{2}\leq r\| \Psi \|_{1}$ for some real number $r$; otherwise, it is unbounded.  In this way, the parameter $\alpha$ plays two roles: it introduces non-Hermiticity, as desired, and results in a metric that is unbounded in the limit of very large $N$. This construction also encompasses other cases, such as $\alpha = 1$ where a treatment has been applied to the $\beta$-Hermite ensemble \cite{marinello2016pseudo,patopseudo}, rendering this framework more general.

In order to investigate the stability of the spectrum of matrices belonging to the $L(\alpha)$ family, we introduce a perturbation through a sparse Hermitian matrix $E$, with elements given explicitly by $E_{ij}=\epsilon \left(\delta_{iN}\delta_{j1}+\delta_{i1}\delta_{jN}\right)$, $\epsilon$ is small. The addition of the corners elements to $L(\alpha)$, $L_{2}(\alpha)=L+E$, explicitly breaks the Hermitian relation produced by $ \eta(\alpha)^{1/2}(L(\alpha)+E)\eta(\alpha)^{-1/2}$, changing the system from an open linear chain to a closed circular chain. Using the new matrix $L_{2}(\alpha,\epsilon)$ and the same $\eta$ operator defined above, we can perform a similarity transformation, $H_{2}(\alpha,\epsilon)=\eta^{1/2}L_{2}(\alpha,\epsilon)\eta^{-1/2}$, to get, 
\begin{equation}
\label{main}
H_{2}(\alpha,\epsilon)=
\left[
\begin{array}{cccc}
	a_{N}^{2}  & a_{N}b_{N-1}  &  & \epsilon \prod\limits_{i=1}^{N-1} (b_{N-i})^{- \alpha}  \\
	a_{N}b_{N-1}  & a_{N-1}^{2}+b_{N-1}^{2} & \ddots  &  \\
	& \ddots  & \ddots  & a_{2}b_{1} \\
	\epsilon \prod\limits_{i=1}^{N-1} (b_{N-i})^{\alpha} &  & a_{2}b_{1} & a_{1}^{2}+b_{1}^{2}
\end{array}
\right].
\end{equation}
 The above pseudo-Hermitian $\beta$-Laguerre ensemble with complex eigenvalues represents the model to be studied in the present work. 

\begin{figure}[hbt!]
	\begin{center}
	\includegraphics[width=0.7\textwidth]{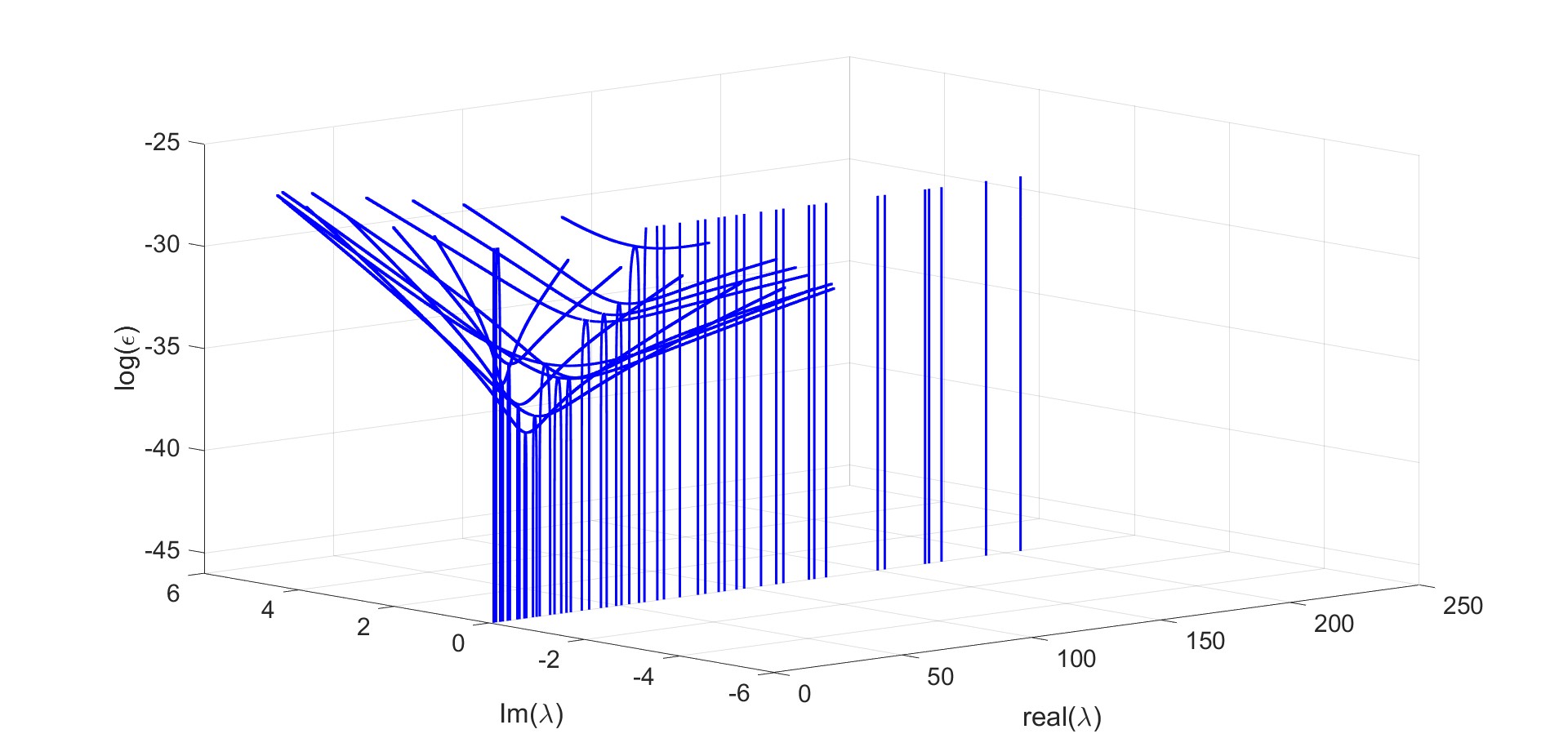}
		\includegraphics[width=0.7\textwidth]{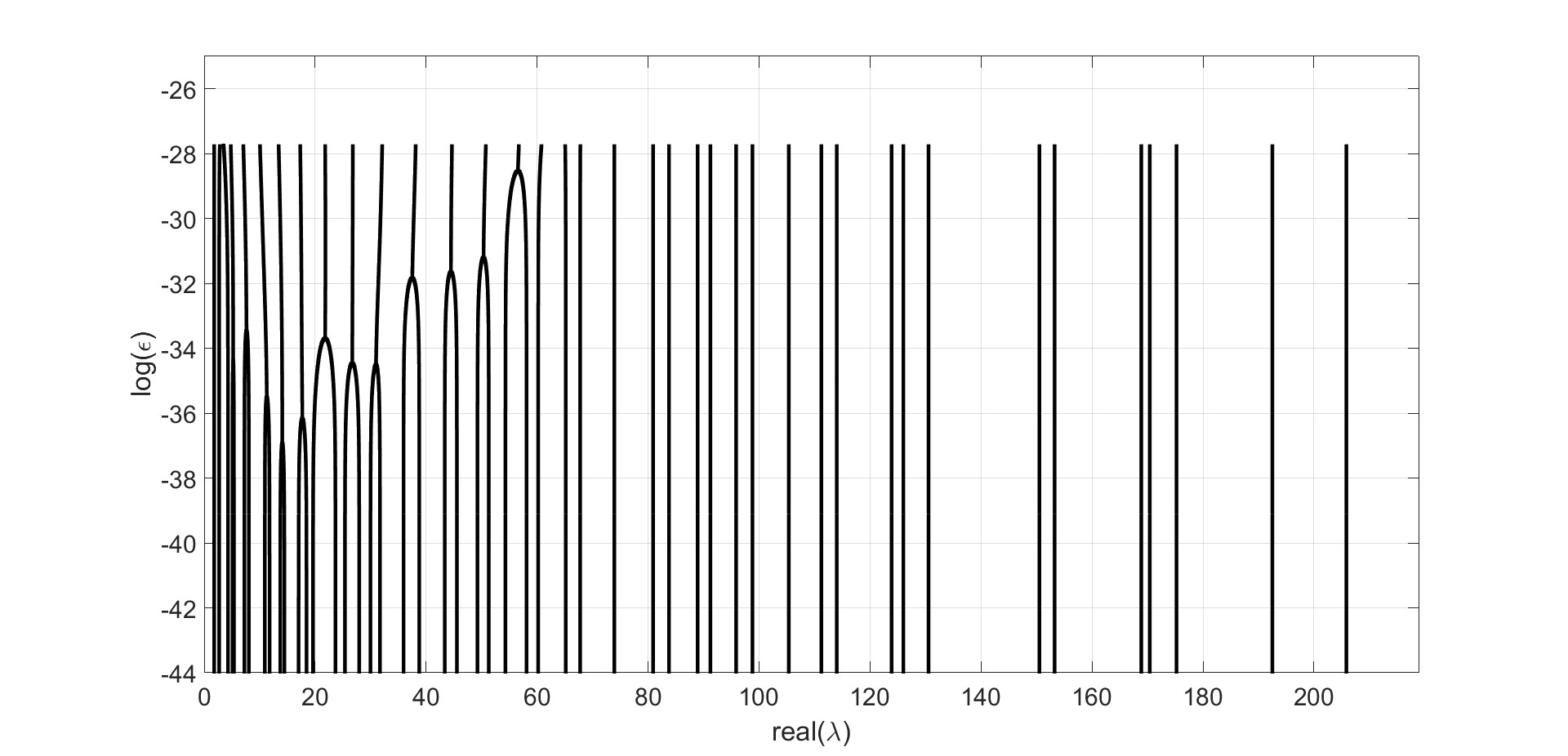} 
			\caption{Eigenvalues of an individual realization of matrix $H_{2}(\alpha,\epsilon)$ with $N=50$, $c=0.75$, and $\beta = 1$ as function of parameter $\epsilon$. Top : Eigenvalues' behavior in the complex ${\rm Re}(\lambda) \times {\rm Im}(\lambda)$ plane. Bottom : Eigenvalues' behavior projected onto the ${\rm Re}(\lambda) \times \epsilon$ plane. As the perturbation parameter increases, the eigenvalues move progressively into the complex domain.} 
	\label{f:1}
	\end{center}
\end{figure}

In order to get some qualitative insight into the behavior of eigenvalues in such an ensemble, we close this Section
by  
illustrating certain features of $H_{2}(\alpha,\epsilon)$ ensemble via numerical simulations, preceding the mathematical analysis which will be presented in the next two Sections. To this end, in Figures \ref{f:1} we depict the behavior of the eigenvalues of a single matrix sample with $N=50, c=0.75$ and $\beta = 1$, as function of the perturbation parameter $\epsilon$. 
 We observe that the lowest eigenvalue of the matrix ensemble remains real; however, coalescing eigenvalues also emerge, which, in physical terms, indicates a symmetry-breaking phenomenon. Real eigenvalues collide and move into the complex plane in conjugate pairs. It is interesting to note the repulsion exhibited by the complex conjugate pair in Figures \ref{f:1}. Such a behavior is similar to the one observed by Bender \textit{et al.} in their investigation of the PT-symmetric Hamiltonians. In general, some amount of eigenvalues remain real for small perturbations, which may be associated with the unbroken PT-symmetry. However, as the perturbation parameter $\epsilon$ increases, more eigenvalues transition into the complex plane.
%The introduction of the perturbation $\epsilon$ encapsulates all non-Hermitian behavior in the corners of what was previously a Hermitian matrix. 
It is also noteworthy that when the eigenvalues of the so-called averaged matrix are computed using the aforementioned parameters, the same behavior as a function of $\epsilon$ is observed. Here, the average matrix is obtained by averaging each element of $B$ before introducing $\alpha$ and $\epsilon$.
\begin{figure}[hbt!]
\begin{center}
		\includegraphics[width=0.7\textwidth]{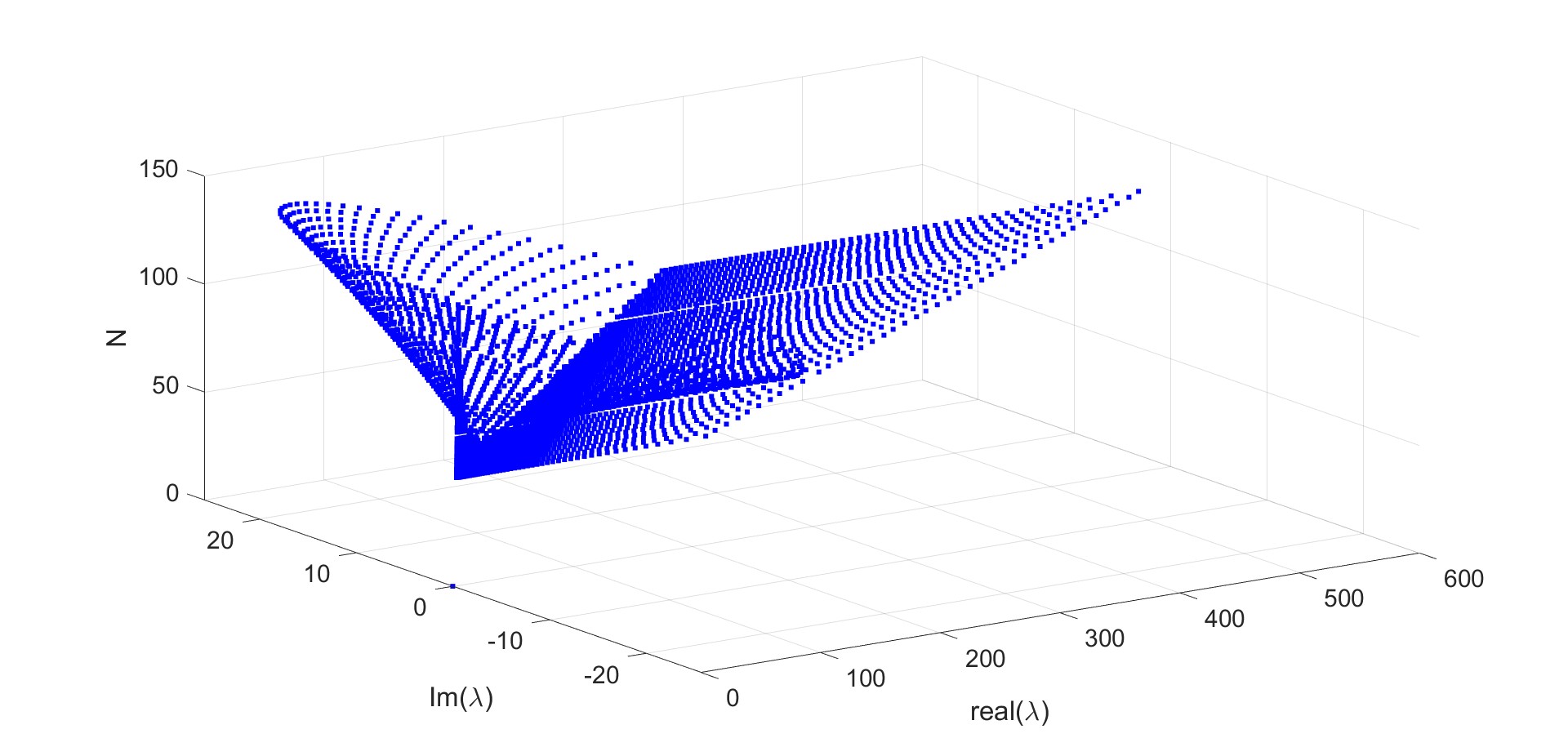}
		\includegraphics[width=0.7\textwidth]{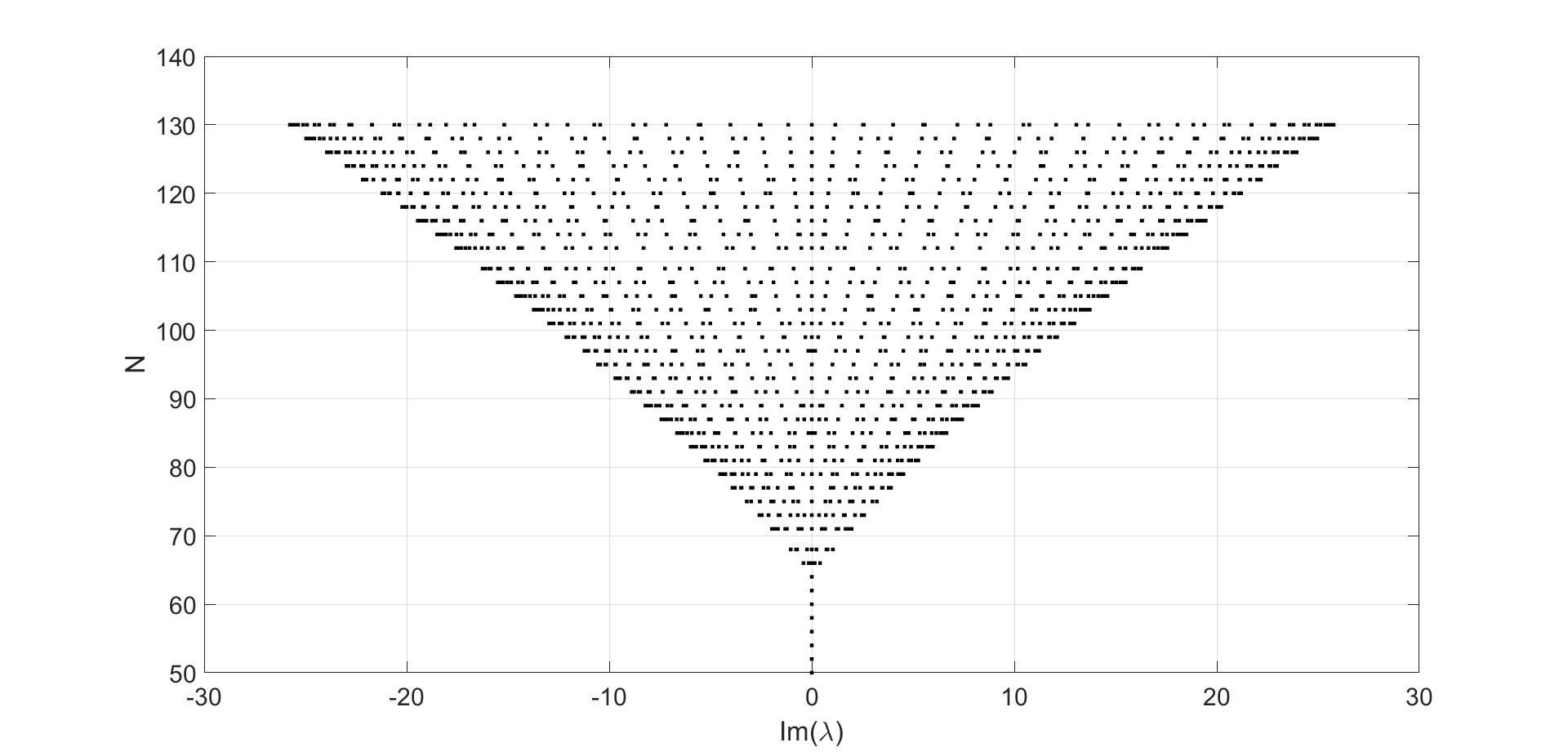} 
	\caption{Eigenvalues in the complex plane for individual realizations of $H_{2}(\alpha,\epsilon)$ as function of matrix dimension $N$, for $\beta = 1$, $c = 0.75$, and $\epsilon = 10^{-10}$. Top : Eigenvalues' behavior in the complex plane. Bottom : Eigenvalues' behavior projected in the ${\rm Im}(\lambda) \times  N$ plane. One observed that the larger system is, more sensitive it becomes to small perturbations.}
	\label{f:2}
	\end{center}
\end{figure}
Figures \ref{f:2}  illustrate the behavior of the eigenvalues as functions of the matrix dimension $N$, for a fixed value of $\epsilon = 10^{-10}$. We observe that for matrices of small dimensions, the eigenvalues always 
 remain real for the perturbation of such a strength. 
 However, as the dimension of the matrix increases, some eigenvalues become complex. This suggests that, with a fixed, arbitrarily small $\epsilon$, a transition to the complex domain occurs as the matrix size increases, as observed also in \cite{khare2000pt}. Larger matrices are more affected by the perturbation parameter. As the size of the matrix increases, the value of the parameter $\epsilon$ at which complex eigenvalues start to appear decreases, which corresponds to the notion of a spontaneous symmetry breaking \cite{van2007spontaneous}. 
 
\section{Mean characteristic polynomial}

 We now aim to determine analytically the loci of the eigenvalues of the ensemble in eq. (\ref{main}), which is standardly done by averaging the resolvent. However, for the model under study 
it seems to be a very complicated or even an unsolvable problem. 
On the other hand, we note  
that in many instances, e.g. for Hermitian matrices, in the large-dimension limit the resolvent starts to exhibit a deterministic behavior, i.e., the system is self-averaging and the resolvent concentrates around a deterministic function \cite{pastur}. Although we cannot prove the self-averaging property of the system in a rigorous way,  
we see some indications in Figures \ref{f:2}
that this may also be the case in the model under study. 
Namely, we observe that the eigenvalues of the average matrix show essentially the same behavior as those of the random one.  Therefore, we will try to circumvent all technical difficulties by first
extracting the information about the loci of the eigenvalues from the mean characteristic polynomial of the matrix in eq. (\ref{main}), and then, in the next Section, validate it via numerical simulations and show that the fluctuations indeed become insignificant in the large-dimension limit.

\subsection{General results for the mean characteristic polynomial of matrices with sufficiently large dimensions}

The characteristic polynomial for $H_{2}(\alpha,0)$ can be obtained by expanding the determinant of the matrix $H_{2}(\alpha,0)- \lambda \mathbb{I}$ by rows and columns, in order to identify a three-term recurrence relation:
\begin{equation}
	P_{N}(\lambda)=(a_{N}^{2}-\lambda)R_{N-1}(\lambda)-(a_{N}b_{N-1})^{2}R_{N-2}(\lambda),
\end{equation}
where the polynomials $R_{N}(\lambda)$ satisfy the recurrence
\begin{equation}
	R_{N}(\lambda)=(a_{N}^{2}+b_{N}^{2}-\lambda)R_{N-1}(\lambda)+(a_{N}b_{N-1})^{2}R_{N-2}(\lambda).
\end{equation}
Averaging the 	above equation, we obtain
\begin{equation}
	\langle P_{N}(\lambda)\rangle=(M\beta -\lambda)\langle R_{N-1}(\lambda) \rangle -M(N-1)\beta^{2}\langle R_{N-2}(\lambda) \rangle,
\end{equation}
and
\begin{eqnarray}
	\langle R_{N-1}(\lambda) \rangle &=& [(M+N-2)\beta - \lambda]\langle R_{N-2}(\lambda)\rangle + \nonumber \\
	&-&(M-1)(N-2)\beta^{2}\langle R_{N-3}(\lambda)\rangle
\end{eqnarray}
 One now notices that these recursion relations are obeyed by two generalized Laguerre polynomials
\begin{equation}
	 \langle R_{N-1}(\lambda)\rangle = (N-1)!\beta^{N-1}L_{N-1}^{M-N+1}\left(\frac{\lambda}{\beta}\right)
\end{equation}
 and 
 \begin{equation}
 	\langle P_{N}(\lambda)\rangle = (N)!\beta^{N}L_{N}^{M-N}\left(\frac{\lambda}{\beta}\right).
 \end{equation}
 We now have all necessary ingredients to define analytically the curve that determines the loci of the eigenvalues
 in the complex domain. 

 Returning to our model in eq. (\ref{main}) 
and assuming that the term $\epsilon \prod\limits_{i=1}^{N-1} (b_{N-i})^{- \alpha}$ can be safely neglected,  which will be validated via numerical simulations \textit{a posteriori}, we  find that the mean characteristic polynomial attains the form
\begin{equation}
	\label{laguerre}
	\langle P_{N}(\lambda) \rangle = N!\beta^{N}L_{N}^{M-N}(\frac{\lambda}{\beta}) = Q,
\end{equation}
where 
\begin{eqnarray}
	Q(\alpha,\epsilon)&=&\epsilon \prod \limits_{i=1}^{N-1} \langle b_{N-i}^{1+\alpha} \rangle \langle a_{N-i+1}\rangle \nonumber\\
	&=&\epsilon \prod \limits_{i=1}^{N-1} \left[ \frac{2^{(1+\alpha/2)}\Gamma(\frac{\nu +1+\alpha}{2})}{\Gamma(\nu/2)}\cdot \frac{\Gamma(\frac{\mu +1}{2})}{\Gamma(\mu/2)} \right],
	\label{originalQ}
\end{eqnarray}
where $\mu=\beta(M-i+1)$ and $\nu=\beta(N-i)$, $i = 1,2, \ldots, N-1$. The associated Laguerre polynomials exhibit oscillatory behavior within the region containing the roots,  and are increasing functions of the argument outside of this region. The amplitude of oscillations is smaller near the origin. Consequently, if $Q$ is sufficiently small  (for given $N$ and $\epsilon$), i.e., less than the oscillations' amplitude, the eigenvalues remain real. On the contrary, upon an increase of $Q$ a pair of complex conjugate eigenvalues appears, as illustrated in the lower panel in Figure \ref{f:1}. The eigenvalue density in the $\beta$-Laguerre ensemble is higher near the smallest eigenvalue, where the Laguerre polynomial's oscillation amplitude is smaller. 
%This observation explains the initial position of the bubble in the spectrum.

\begin{figure}[hbt!]
	\begin{center}
		\includegraphics[width=0.7\textwidth]{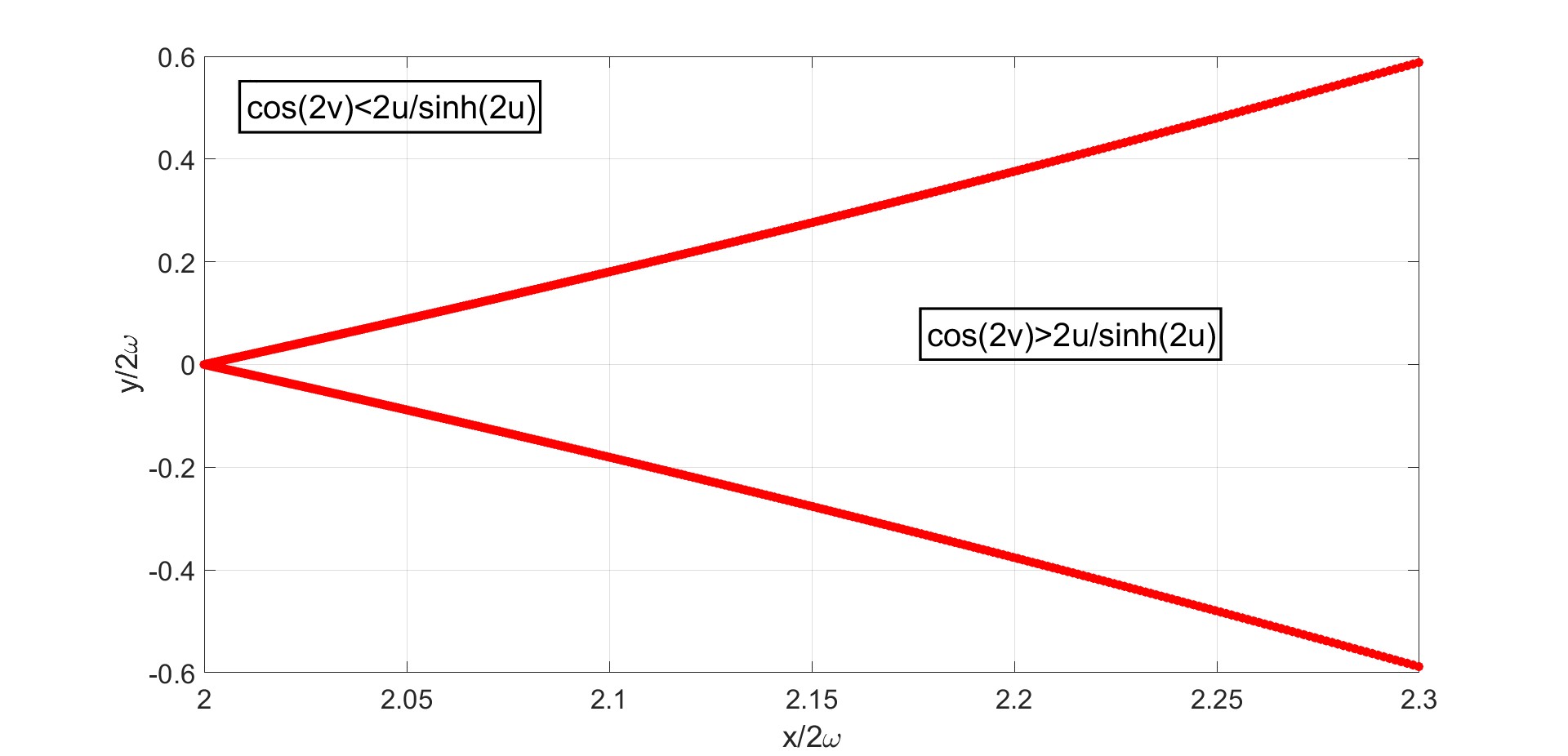}
		\includegraphics[width=0.7\textwidth]{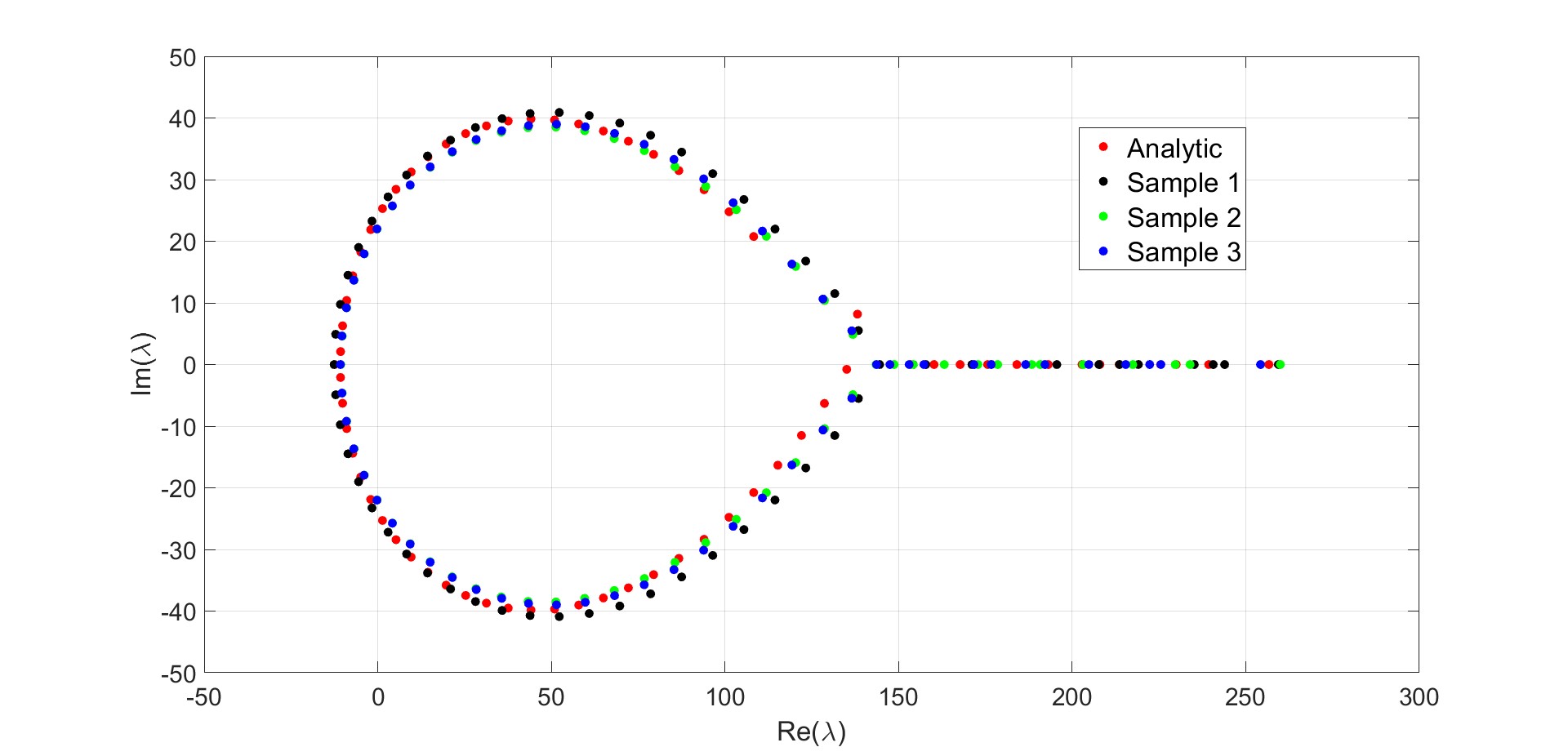} 
		\caption{Eigenvalues of the $H_{2}(\alpha,\epsilon)$ 
			ensemble with the parameters $N=60$, $M=80$, $\beta=1$, $\alpha=0.5$ and $\epsilon=10$. Top: Stokes lines (see eqs (\ref{Xstokes}) and (\ref{Ystokes})) on the complex plane. Bottom: Balloon-like structure formed by the eigenvalues on the complex plane. This structure has an apparent cusp at its right extremity which is followed by a discrete finite line of real eigenvalues.
			Red filled circles - analytical solution of the coupled eqs. (\ref{U}) and (\ref{V}). Filled colored circles - numerically evaluated eigenvalues of three random samples.}
		\label{f:3} % <-- agora está dentro e depois do \caption
	\end{center}
\end{figure}

 Denoting next  $z=(4N+2\delta+2)\cos^{2}\phi$ and assuming that $z$ is large, we take advantage 
of the following asymptotic formula for the Laguerre polynomials \cite{szeg1939orthogonal},
\begin{eqnarray}
	\label{expansionlaguerre}
	L_{N}^{\delta}(z)&=(-1)^{N}(\pi \sin\phi)^{-1/2}e^{z/2}z^{-\delta/2 -1/4}N^{\delta/2 -1/4}\nonumber\\
	&\times \left[ \sin\{\omega(\sin(2\phi)-2\phi) +3\pi/4 \} \right] ,
\end{eqnarray}
 which can be extend to the complex domain by setting $\phi=v-iu$, where  
the components of the complex number $z=x+i y$ are expressed through $v$ and $u$, (with $u>0$), as
\begin{equation}
	\label{x}
	x=(M+N+1)\{1+ \cosh 2u \cos 2v \},
\end{equation}
and
\begin{equation}
	\label{y}
	y=(M+N+1)\sinh 2u \sin 2v.
\end{equation}
Next, we have from eq. (\ref{expansionlaguerre}),
\begin{eqnarray*}
	\sin(2\phi)-2\phi +3\pi/4 &= \sin(2v)\cosh(2u)-2v+\frac{3\pi}{4\omega}\\
	&\times i\left[-\sinh(2u)\cos(2v)+2u\right],
\end{eqnarray*}
where $\omega = N+(M-N+1)2$ depends only on the dimensions of the matrix. 
From the expression in the brackets in the second line in the latter equation we infer that the equation
\begin{equation}
	2u-\sinh(2u)\cos(2v)=0,
\end{equation}
defines the Stokes lines in the complex plane.  These Stokes lines are 
depicted in the upper panel in Figure \ref{f:3} and are implicitly defined by the parametric equations
\begin{equation}
	\label{Xstokes}
	x=(M+N+1)\left[ 1+\frac{2u}{\tanh(2u)} \right],
\end{equation}
and
\begin{equation}
	\label{Ystokes}
	y=\pm (M+N+1)\sqrt{(\sinh(2u))^{2}-(2u)^{2}}
\end{equation}    
 Within a triangular area delimited by the Stokes lines, we require that $\cos(2v)\sinh(2u)-2u<0$ in order to ensure that  in the limit $u \rightarrow 0$ the condition $0<\phi<\pi/2$ for the validity of the asymptotic expansion is satisfied.  
Under the latter condition, in the limit of large $N$ we can use the following approximation $\cosh(\xi)\approx \sinh(\xi)\approx e^{\xi}/2$ for hyperbolic functions whose arguments lie in the positive region within the Stokes lines in the expression bellow
\begin{eqnarray*}
	&\sin&\left( \omega\left[\sin(2\phi) -2\phi+\frac{3\pi}{4\omega}\right] \right)= \nonumber \\
	&\sin&(\omega \sin(2v)\cosh(2u)-2v+\frac{3\pi}{4\omega})\cosh(\omega \sinh(2u)\cos(2v)-2u ) \nonumber \\
	&-&i\sinh(\omega \sinh(2u)\cos(2v)-2u)\cos(\omega \sin(2v)\cosh(2u)-2v+\frac{3\pi}{4\omega}).
\end{eqnarray*}
Therefore, the hyperbolic functions can be replaced by exponentials, which entails
\begin{eqnarray}
\label{m}
	L_{N}^{\delta}(z)&=& \exp\left[\omega \left(\Omega + 1+2u+\cos(2v)\exp^{-2u} +i2v-i\sin(2v)\exp^{-2u} \right) \right], \nonumber\\
\Omega &=&\frac{1}{\omega}\left[\frac{2(M-N)+1}{4}\log(Nz)-\log(\pi\sin\phi)^{2}\right]	 \,.
\end{eqnarray} 
Note that in the limit $N\rightarrow \infty$ the value of the parameter $\Omega$ becomes negligibly small.

Substituting the asymptotic expression (\ref{m}) into eq. (\ref{laguerre}) and taking the large $N$ limit, we derive the following pair of coupled equations: 
\begin{equation}
	\label{U}
	1+2u+\exp(-2u)\cos(2v)=\frac{1}{\omega}\ln\left( \frac{Q(\alpha,\epsilon)}{N!\beta^{N}} \right),
\end{equation} 
and
\begin{equation}
	\label{V}
	2v-sin(2v)\exp(-2u)=\frac{(2k+M-N+1)}{M+N+1}\pi,
\end{equation}
which determine the coordinates $u$ and $v$ of the eigenvalues on the complex plane. Note that $k$ is an integer ranging from $1$ to $N$, controlling the variation of the angle between $0$ and $\pi$. Equations (\ref{U}) and (\ref{V}) can be solved iteratively. The analytical solution of eqs. (\ref{U}) and (\ref{V}) (red closed circles) is presented in the lower panel in  Figure \ref{f:3}, together with the numerically evaluated eigenvalues of three random samples from the $H_{2}(\alpha,\epsilon)$ ensemble, using the parameters $N = 60$, $M = 80$, $\beta = 1$, $\alpha = 0.5$, and $\epsilon = 10$. Numerically evaluated eigenvalues are depicted by colored filled circles (see the inset in the lower panel in Figure \ref{f:3}).

Capitalizing on our eqs. (\ref{U}) and (\ref{V}),  as well as on the results of the numerical evaluation of the eigenvalues of random sample matrices, 
	we conclude  that the eigenvalues form a closed balloon-like structure on the complex plane (see Figure \ref{f:3}, lower panel) which is followed from its right by a finite line of real eigenvalues. No real eigenvalue appears from the left of the balloon-like structure. 
	We also observe that the numerically evaluated eigenvalues
	indeed seem to concentrate around the deterministic analytical curve defined  
	by eqs. (\ref{U}) and (\ref{V}), which are deduced from the mean characteristic polynomial.  
	However, some fluctuations are apparent and at the moment we are not in position to say if the fluctuations will become more significant if we increase the dimension, or, on the contrary, their relative magnitude will become less important. This is the question which we will address in the next Section. 
	
	A legitimate question is also how the parameter $\beta$, which controls, e.g.,  the mutual repulsion of eigenvalues,  may affect the behavior. In quest for the answer, we produced a similar analysis as in Figure \ref{f:3} for several other values of $\beta$. We observed that, not counter-intuitively,  upon lowering $\beta$ the spread of fluctuations becomes more significant. On the contrary,  increasing $\beta$ we increase the repulsion of the eigenvalues which leads to their "crystallization"  on the curve defined by eqs. (\ref{U}) and (\ref{V}) and, in the consequence, to a suppression of the fluctuations, both along this curve and in the perpendicular to it direction. In the next Subsection, we will reconsider the dependence on $\beta$ in the asymptotic limit when the dimensions tend to infinity.  
	
Lastly, we note
that although the analytical eqs. (\ref{U}) and (\ref{V}) are derived under the assumption that the dimensions are sufficiently large, 
the random sample matrices which we used to produce the lower panel in Figure \ref{f:3} are not that large, showing nonetheless a very nice agreement between the theoretical predictions and the numerically evaluated eigenvalues. 
In fact, the choice of such modest dimensions is certainly a limitation and
is due to a sensitivity 
 introduced by the sparse perturbative matrix $E$, which renders the spectrum more
 susceptible to a misinterpretation, especially in the asymptotic limit. Such a misinterpretation can be understood by considering the effect of the perturbative matrix $E$ on the spectrum of the unperturbed matrix $H(\alpha)$, i.e., by evaluating the shift $\|\lambda_{0} - \lambda\|$ between the spectra of the unperturbed and perturbed matrices.
 
	The misinterpretations of the spectra can be characterized in various ways. One approach to studying the spectral sensitivity of the non-Hermitian matrices is through the so-called $\epsilon$-pseudospectrum. According to the definition presented in \cite{reichel1992eigenvalues} (see also \cite{trefethen2020spectra} for a general review), given some $\epsilon > 0$, a number $\lambda$ is an $\epsilon$-pseudo-eigenvalue of $H_2(\epsilon, \alpha)$ if,
	\begin{equation}
		\| (\lambda  \mathbb{I}-H_{2}(\epsilon,\alpha))^{-1} \| \geq \epsilon^{-1},
	\end{equation}
	where $\| \bullet \|$ denotes the $2$-norm.  For such a definition, the matrix represents the resolvent of $H_{2}(\epsilon,\alpha)$ at $\lambda$. In simple terms, as mentioned above, the pseudospectrum helps to understand the relation,
	\begin{equation}
		\| \lambda_{0}-\lambda \| \leq \kappa_{\lambda}\epsilon+O(\epsilon^{2}),
	\end{equation}
	where $\kappa_\lambda$ denotes the condition number, which for an appropriately defined metric 
	suitable for a problem at hand, allows us to determine the effect of perturbations in non-normal matrices, i.e., matrices $M$ for which $[M, M^\dagger] \neq 0$.
	In physical terms, this implies that, for a series of individual realizations of $H_2(\epsilon, \alpha)$ with large $N$ and a bounded spectrum, the non-equivalence
	\begin{equation}
		\lim\limits_{N\rightarrow \infty} \lim\limits_{\epsilon \rightarrow 0} \Lambda_{\epsilon}(H_{2}(\epsilon,\alpha)) \neq \lim\limits_{\epsilon \rightarrow 0} \lim\limits_{N\rightarrow \infty} \Lambda_{\epsilon}(H_{2}(\epsilon,\alpha))
	\end{equation}
	where $\Lambda_\epsilon(H_2(\epsilon, \alpha))$ denotes the $\epsilon$-pseudospectrum, is reminiscent of a symmetry breaking. The order in which the thermodynamic limit ($N\rightarrow \infty$) and the zero-pertubation limit ($\epsilon \rightarrow 0$) are taken is important. To avoid such complications, which are beyond the scope of our work, the above chosen values of parameters were used.

	\subsection[Asymptotic Behavior]{Behavior in the asymptotic limit $N \to \infty$}

 Equations (\ref{U}) and (\ref{V}) are valid when either $\epsilon$ is sufficiently small or $N$ is sufficiently large.  
We turn to the asymptotic limit $N \to \infty$, in which  eqs. (\ref{U}) and (\ref{V}) considerably simplify permitting to get an insight into many aspects of the behavior.  At the end of this Subsection we will also briefly comment  on some peculiarities of the limit $\epsilon \to 0$. 
For fixed small $\epsilon$ and $N \to \infty$, we have, taking advantage of the Stirling's approximation, that in the leading in this limit order  
\begin{equation}
	\label{logN}
	\frac{1}{\omega}\log\left(\frac{Q(\alpha,\epsilon)}{N!\beta^{N}}\right)\simeq \frac{\alpha}{2}\log(N), 
\end{equation}
where the symbol $\simeq$ signifies that we consider the leading in the limit $N \to \infty $ behavior.  Interestingly enough, in this limit the leading behavior appears to be independent of $\beta$, 
but is linearly dependent on the parameter $\alpha$. The logarithmic divergence of the right-hand-side of eq. (\ref{logN}) implies that $u$ defined by eq. (\ref{U}) also tends to infinity as $N \to \infty$. As a consequence, the function $\exp(-2u)$ vanishes, and 
eqs. (\ref{U}) and (\ref{V}) decouple from each other. The solutions for $u$ and $v$ are then given by
\begin{eqnarray}
	\label{asolution}
	2u&\approx& \frac{\alpha}{2}\log N\\
	2v&\approx& \frac{2k}{M+N}\pi.
\end{eqnarray}
such that the coordinates of complex eigenvalues obey
\begin{equation}
	\label{circle}
	(x-N)\approx \frac{N e^{2u}}{2}\cos(2v),
\end{equation}
and
\begin{equation}
	y\approx \frac{N e^{2u}}{2}\sin(2v)
\end{equation}

\begin{figure}[hbt!]
	\begin{center}
		\includegraphics[width=0.7\textwidth]{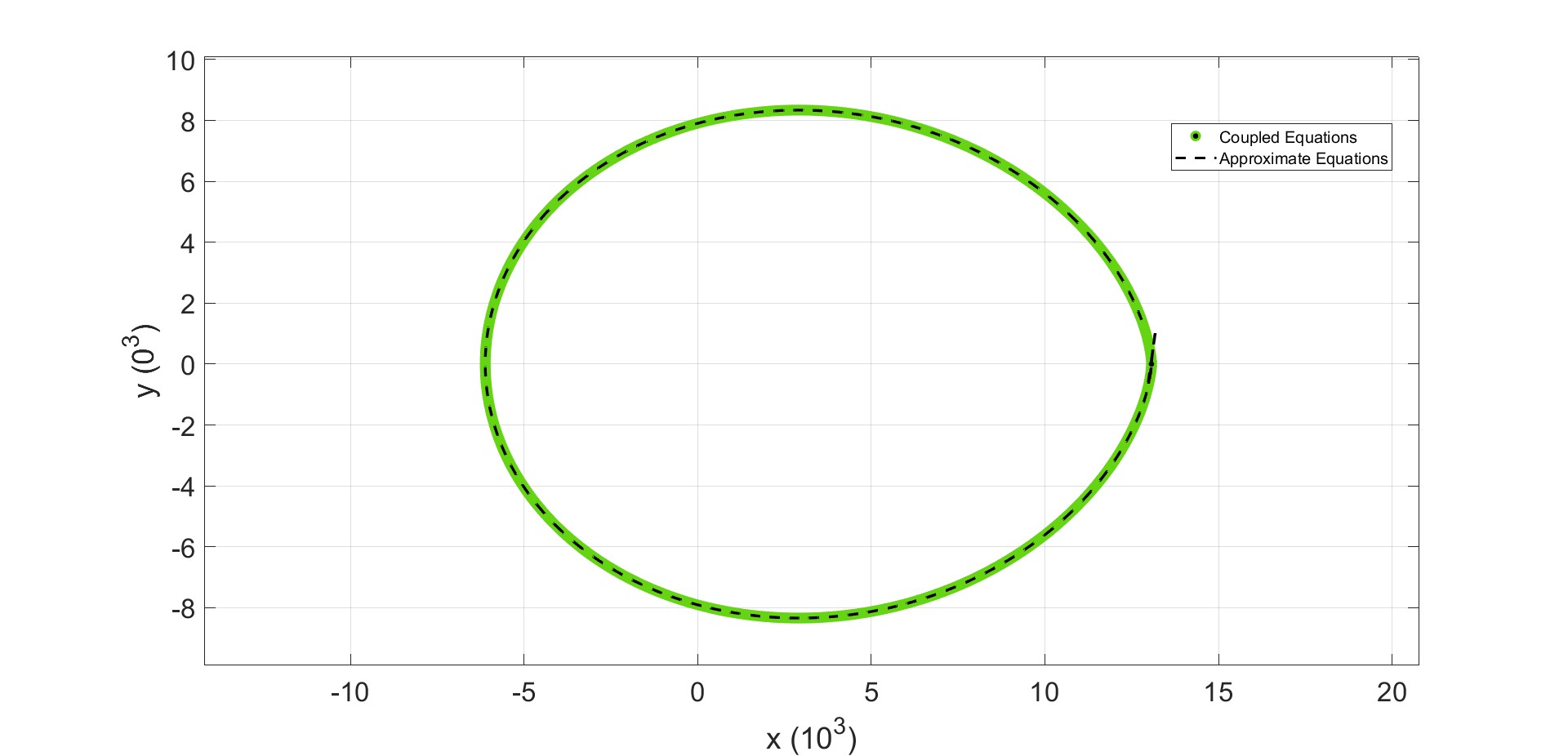}
		\includegraphics[width=0.7\textwidth]{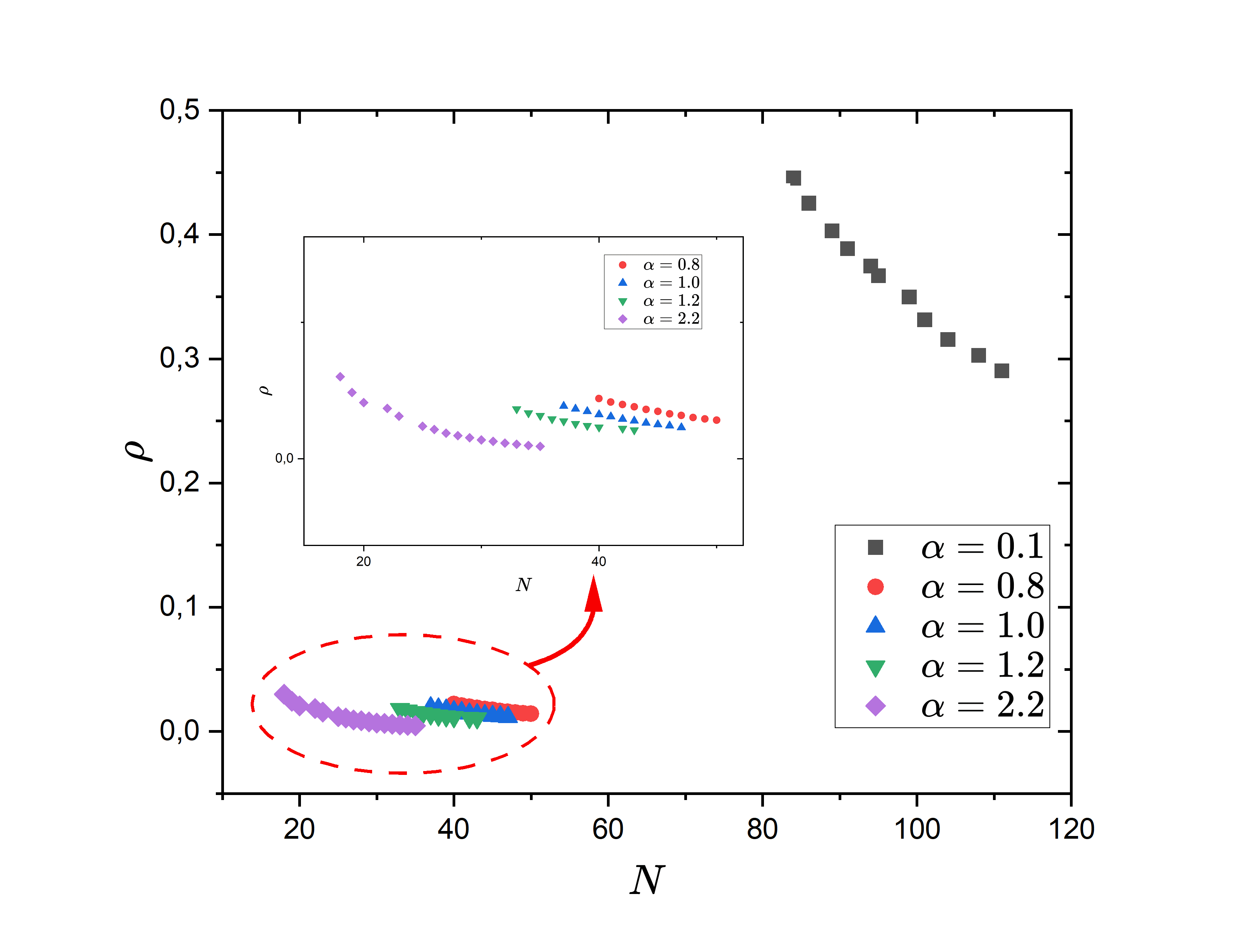} 
		\caption{Limiting behavior for $N \to \infty$. Top: Solution of the coupled eqs.  (\ref{U}) and (\ref{V}) is compared against the limiting form in eqs. (\ref{circle}) for $N = 30000$, $c = 0.85$, $\beta = 1$, and $\alpha = 0.5$. The agreement is very good except for a narrow vicinity of the cusp at the right extremity of the balloon-like structure. Bottom: The density of eigenvalues on the circumference of the balloon-like structure as function of $N$ for different values of $\alpha$.  To facilitate the interpretation of the region delineated by the dashed line, we plotted the graph indicated by the red arrow. The inset graph displays the curves corresponding to $\alpha \geq 0.8$.}
		\label{f:4} % <-- corrigido: agora está dentro do caption block
	\end{center}
\end{figure}

 This signifies that in this limit the ballon-like structure with a cusp at its right extremity should attain a limiting shape which is a perfect circle centered at $(N,0)$ and having the radius $R = \left(N \exp(2u)/2\right) \simeq N^{1+ \alpha/2}/2$. In the upper panel in  Figure \ref{f:4}  we present a comparison of the solution of the coupled eqs. (\ref{U}) and (\ref{V}) for $N=50000$, $c=0.85$, $\alpha=0.5$ and  $\beta=1$,  and the limit solution defined in  eq. (\ref{circle}). Overall, we observe a perfect agreement between the two solutions, except for a narrow vicinity of the cusp. Apparently, this singularity will always be present for any finite $N$ and disappear only in the formal limit $N \to \infty$. Note that on this panel in Figure \ref{f:4} we present only the complex eigenvalues and skip the real ones.

 An interesting question is how densely the eigenvalues are populating the perimeter of the balloon-like structure. Given that the number of eigenvalues is proportional to $N$, while the circumference $\gamma$ of the balloon-like structure is  proportional to $R \simeq N^{1+ \alpha/2}$, we have that the density of eigenvalues on the circumference behaves in the limit $N \to \infty$ as
\begin{equation}
	\rho(N)=\frac{N}{\gamma}\propto \frac{1}{N^{\alpha/2}}.
\end{equation}
	Here, it is important to make an observation regarding the value of $\alpha$. In eq. \ref{main}, we neglect the top-right corner element $(1, N)$ in the limit of large $N$. Under this assumption, the value of $\alpha$ cannot be zero, since this would produce a Hermitian matrix with entirely real eigenvalues. This is the condition under which the above equation remains valid. The density $\rho$ vanishes in the limit $N \to \infty$ for any $\alpha > 0$, implying that the arc length between neighboring eigenvalues grows without bound as $N$ increases. This behavior is evident in the lower panel of Figure \ref{f:4}, where we plot the eigenvalue \textit{density} as a function of $N$ for different values of $\alpha$. Moreover, we observe that the difference in arc length between neighboring eigenvalues depends on the positions of those eigenvalues. 

 Lastly, we briefly comment on the limit $\epsilon \to 0$, which should be indeed taken with an appropriate care. 
We focus on the asymptotic behavior of $Q(\alpha,\epsilon)$ as $\log(\epsilon)\rightarrow -\infty$ for $\epsilon \rightarrow 0$. To analyze the passage to the unperturbed Laguerre ensemble, we must first note that $N$ cannot be kept fixed if we want to obtain a consistent result. This is due to a discontinuity in the transition from the real line to the complex plane: as $\epsilon$ decreases arbitrarily, $N$ must be  increased accordingly.
For large $N$, and initially fixed $\epsilon$, using Stirling's approximation, we can write:
\begin{equation}
	\frac{1}{\omega}\log\left(\frac{Q(\alpha,\epsilon)}{N!\beta^{N}}\right)\approx \frac{\log(\epsilon)}{N}+\frac{\alpha}{2}\log(N).
\end{equation}
The unperturbed matrix corresponds to the case where $Q=0$ in eq. (\ref{laguerre}). We observe that, if $N$ is kept fixed, the original behavior of the unperturbed matrix cannot be recovered due to the divergence of the logarithmic term. One way to address this issue is to introduce a cutoff in order to regularize the divergence. It is important to note that this regularization is not unique. Nevertheless, by choosing $\epsilon = 1/N$, we obtain a 
consistent result, since, evidently
\begin{equation}
	\lim_{N\rightarrow \infty} \frac{\log(\frac{1}{N})}{N}=0 .
\end{equation}

\section{Variance of the Characteristic Polynomial}

\begin{figure}[hbt!]
	\begin{center}
		\includegraphics[width=0.7\linewidth]{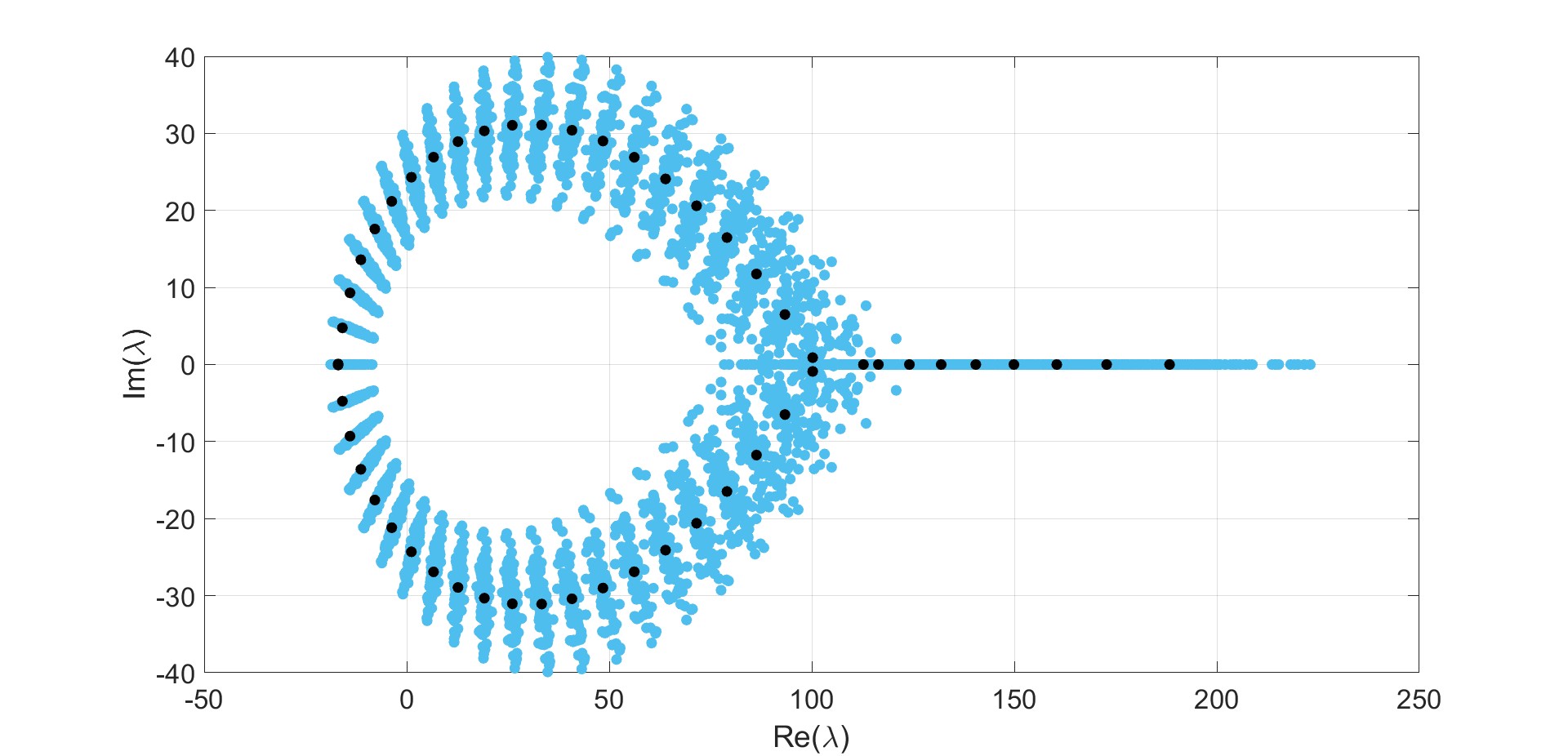}
		\caption{ Fluctuations around the roots of the mean characteristic polynomial. The blue filled circles depict the spread of fluctuations 
	 	of the eigenvalues of $100$ samples of matrices with  $N=50$, $M=54$,$\beta=1$,$\alpha=1$, and $\epsilon=10^{-10}$. The black filled circles represent the numerical solution of eqs.  (\ref{U}) and (\ref{V}).}
	 	\label{f:5}
	\end{center}
\end{figure}

 The above results on the loci of complex eigenvalues were deduced from the mean characteristic polynomial 
and also seemingly agree quite well with the behavior observed in numerical simulations. On the other hand, there is no guarantee that this behavior will persist for an arbitrary number of ensemble members, even in the large-$N$ limit. In this section, we revisit this question and focus on the spread of fluctuations around the mean values.
Figure \ref{f:5} shows the distribution of eigenvalues for $100$ realizations of the matrices $H_2(\alpha,\epsilon)$  in the complex plane (blue filled circles), compared with the solution of the eqs. (\ref{U}) and (\ref{V}), indicated by black filled circles. A salient feature is that the fluctuations are anisotropic. They are strongly suppressed along the circumference of the balloon-like structure, where distinct gaps between the eigenvalues are visible, but become significantly more pronounced in the direction perpendicular to this boundary. Additionally, the magnitude of the fluctuations varies along the structure, being notably smaller at the left extremity and increasing substantially toward the right extremity.

Consider the first two moments of the above-defined characteristic polynomial of a single matrix realization in the ensemble $H_{2}(\alpha, \epsilon)$:
\begin{equation}
	\label{M1}
	M_{1}(z)=\langle \det[H_{2}(\alpha, \epsilon)-z\mathbb{I}] \rangle,
\end{equation}
and
\begin{equation}
	\label{M2}
	M_{2}(z,w)=\langle \det[(H_{2}(\alpha, \epsilon)-z\mathbb{I})\cdot(H_{2}(\alpha, \epsilon)^{\dagger}-w\mathbb{I})] \rangle,
\end{equation}
 here, the angle brackets denote averaging over an ensemble of matrices; "$\dagger$"' represents the composition of the transpose and complex conjugation operations (for real matrices, this operation reduces to the transpose alone); $\mathbb{I}$ denotes the appropriate identity matrix; and the bar indicates complex conjugation.
Knowing these two moments will allows us to quantify the fluctuations by analyzing the variance
\begin{equation}
	\label{variance}
	\Sigma(z)=M_{2}(z,\bar{z})-\overline{M_{1}(z)}M_{1}(z),
\end{equation}
and the relative variance 
\begin{equation}
	\label{relative}
	\Sigma_{r}(z)=\frac{M_{2}(z,\bar{z})}{\overline{M_{1}(z)}M_{1}(z)}-1 ,
\end{equation}
where the latter provides a standard criterion for determining whether a disordered system exhibits a self-averaging behavior or not \cite{pastur}.

 A sample matrix drawn from our pseudo-Hermitian ensemble, despite having complex eigenvalues, has zero probability of exhibiting eigenvalue degeneracy. This follows from the continuous probability distributions defined in eqs. (\ref{disF}) and (\ref{PB}). It is important to note that averaging the elements of the matrix $H_2(\alpha,\epsilon)$ yields a deterministic matrix, which is no longer random. Denoting the eigenvalues of this deterministic matrix as $\mu_1, \mu_2, ..., \mu_N$, we can rewrite eq. (\ref{M1}) in the following form:
\begin{equation}
		M_{1}(z) =\prod\limits_{i=0}^{N}\left(\mu_{i}-z\right) = z^{N}+\left(\sum_{i=1}^{N} \mu_{i}\right)z^{N-1}+O(z^{N-2}),
\end{equation}
and, squaring the above equation, we have
\begin{equation}
	\label{M1}
		M_{1}(z)\overline{M_{1}(z)}=|z|^{2N}+\gamma|z|^{2N-2}\left[\gamma+z+\bar{z}\right]+O(|z|^{2N-4}),
\end{equation}
where $\gamma=\left(\sum_{i=1}^{N} \mu_{i}\right)$. 

The second moment of the characteristic polynomial can be expressed as
\begin{eqnarray}
	\label{newM2}
	M_{2}(z,w) &=& \langle \det[(H_{2}(\alpha, \epsilon)-z\mathbb{I})\cdot(H_{2}(\alpha, \epsilon)^{\dagger}-w\mathbb{I})] \rangle \nonumber \\
	&=& \langle \det[(H_{2}(\alpha, \epsilon)-z\mathbb{I})\cdot(\eta H_{2}(\alpha, \epsilon) \eta^{-1}-w\mathbb{I})] \rangle \nonumber \\
	&=& \sum \limits_{i=0,j=0}^{N,N} \langle f_{N}^{(i)} f_{N}^{(j)} \rangle (-z)^{N-i} (-w)^{N-j}.
\end{eqnarray}
where in the expression in the second line we take advantage of the pseudo-Hermitian property of the ensemble $H_2(\alpha,\epsilon)$. In eq. (\ref{newM2}), the coefficients cannot be obtained directly from the roots of the characteristic polynomial. They are symmetric functions of the $N$ eigenvalues of each sample matrix $S$, such that
\begin{eqnarray*}
	f_{N}^{0} &=& 1 \\
	f_{N}^{1} &=& \lambda_{1}+\lambda_{2}+\dots+\lambda_{N}	\\
	f_{N}^{2} &=& \lambda_{1}\lambda_{2}+\lambda_{1}\lambda_{3}+\dots+\lambda_{N-1} \lambda_{N} \\
	,&\dots&,\\
	f_{N}^{N} &=& \lambda_{1}\lambda_{2}\dots\lambda_{N}.
\end{eqnarray*}
Each $f_{N}^{i}$ is the sum  of all unique permutations of the $i-{\rm uple}$ products of the $\lambda_{1},\lambda_{2},\dots,\lambda_{N}$. The coefficients of order $N$ can be obtained from lower-order coefficients  using the recursion
\begin{equation}
	f_{N}^{(i)}=f_{N-1}^{(i)}+\lambda_{N}f_{N-1}^{(i-1)},
\end{equation}
which  holds for all $i$ as long as we impose the boundary conditions, 
\begin{equation*}
	f_{N}^{0}=1, f_{N}^{i}=0,\quad i>N.
\end{equation*}
 One notices that $M_{1}(z)\overline{M_{1}(z)}$ and $M_{2}(z,\bar{z})$ are, in  fact, the monic polynomials; that being,  non-zero univariate polynomials in which the leading coefficient (the non-zero coefficient of the highest degree) is equal to $1$. For instance, we can explicitly write the first few higher-degree terms of $M_{2}(z,\bar{z})$ as
\begin{equation}
	\label{M2}
M_{2}(z,\bar{z})= |z|^{2N}+\langle f_{N}^{1}f_{N}^{1} \rangle |z|^{N-2}\left[\langle f_{N}^{1}f_{N}^{1} \rangle +z+\bar{z}\right]+O(|z|^{N-4}) 
\end{equation}
Combining eqs. (\ref{M1}) and (\ref{M2}), we can write down eq. (\ref{variance}) as
\begin{equation}
	\Sigma(z)=|z|^{2N-2}\left[\langle f_{N}^{1}f_{N}^{1} \rangle^{2}-\gamma^{2}+(z+\bar{z})\left(\langle f_{N}^{1}f_{N}^{1} \rangle -\gamma\right)\right]+O(|z|^{2N-4}),
\end{equation}
and rewrite eq. (\ref{variance}) as
\begin{equation}
	\label{varapp}
	\Sigma_{r}(z)\approx \frac{|z|^{2N-2}\left[\langle f_{N}^{1}f_{N}^{1} \rangle^{2}-\gamma^{2}+(z+\bar{z})\left(\langle f_{N}^{1}f_{N}^{1} \rangle -\gamma\right)\right]}{|z|^{2N}+\gamma|z|^{2N-2}\left[\gamma+z+\bar{z}\right]}.
\end{equation}

 From the previous sections and Figure (\ref{f:1}), we note that the transition of eigenvalues from the real line to the complex plane does not significantly shift the real part of the eigenvalues. That is, the real parts of the conjugate pairs that move into the complex plane remain close to the values of the original real eigenvalues. We can use the lower bound of the Marchenko-Pastur distribution to establish the inequality $| \lambda_{m}| \geq \lambda_{-}$ where $\lambda_{-} = N\beta \left( \sqrt{M/N} - 1 \right)^2$, and $\lambda_{m}$ is the rightmost eigenvalue in the complex plane distribution.
For large $N$, if eq. (\ref{varapp}) is evaluated at $z = \lambda_{-}$, we observe that the relative variance in this regime is zero. Furthermore, increasing $N$ shows that the relative variance for $z$ near or greater than $\lambda_{-}$ remains zero:
\begin{equation}
	\lim \limits_{N\rightarrow \infty} \Sigma_{r}(|z| \geq \lambda_{-}) =0.
\end{equation}
Equations (\ref{x}) and (\ref{y}) indicate that both the real and imaginary parts of the eigenvalues increase with $N$. This observation supports the conclusion that in the vicinity of the eigenvalues, the relative variance vanishes—i.e., the system exhibits self-averaging behavior.

\begin{figure}[hbt!]
	\begin{center}
		\includegraphics[width=0.7\textwidth]{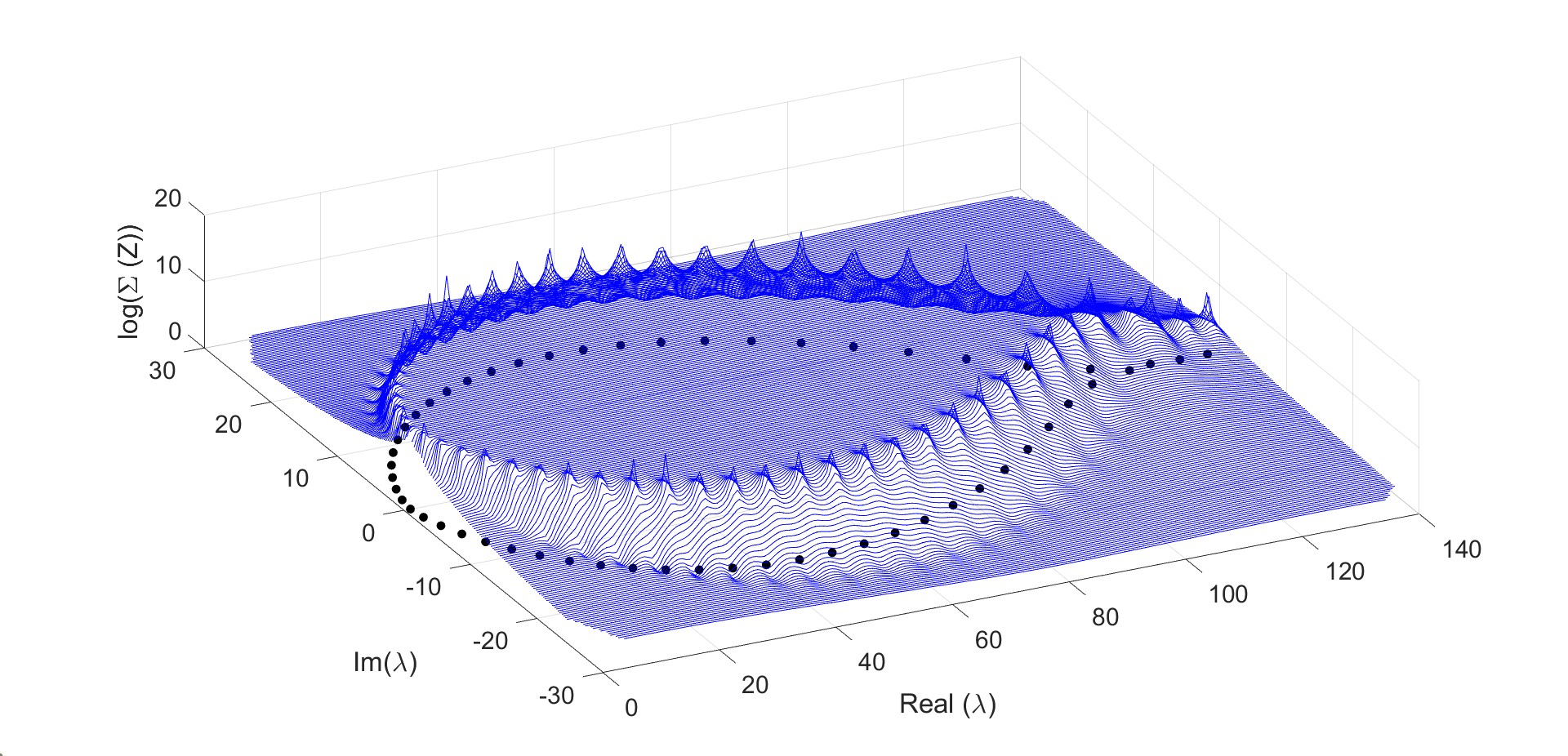}
		\includegraphics[width=0.7\textwidth]{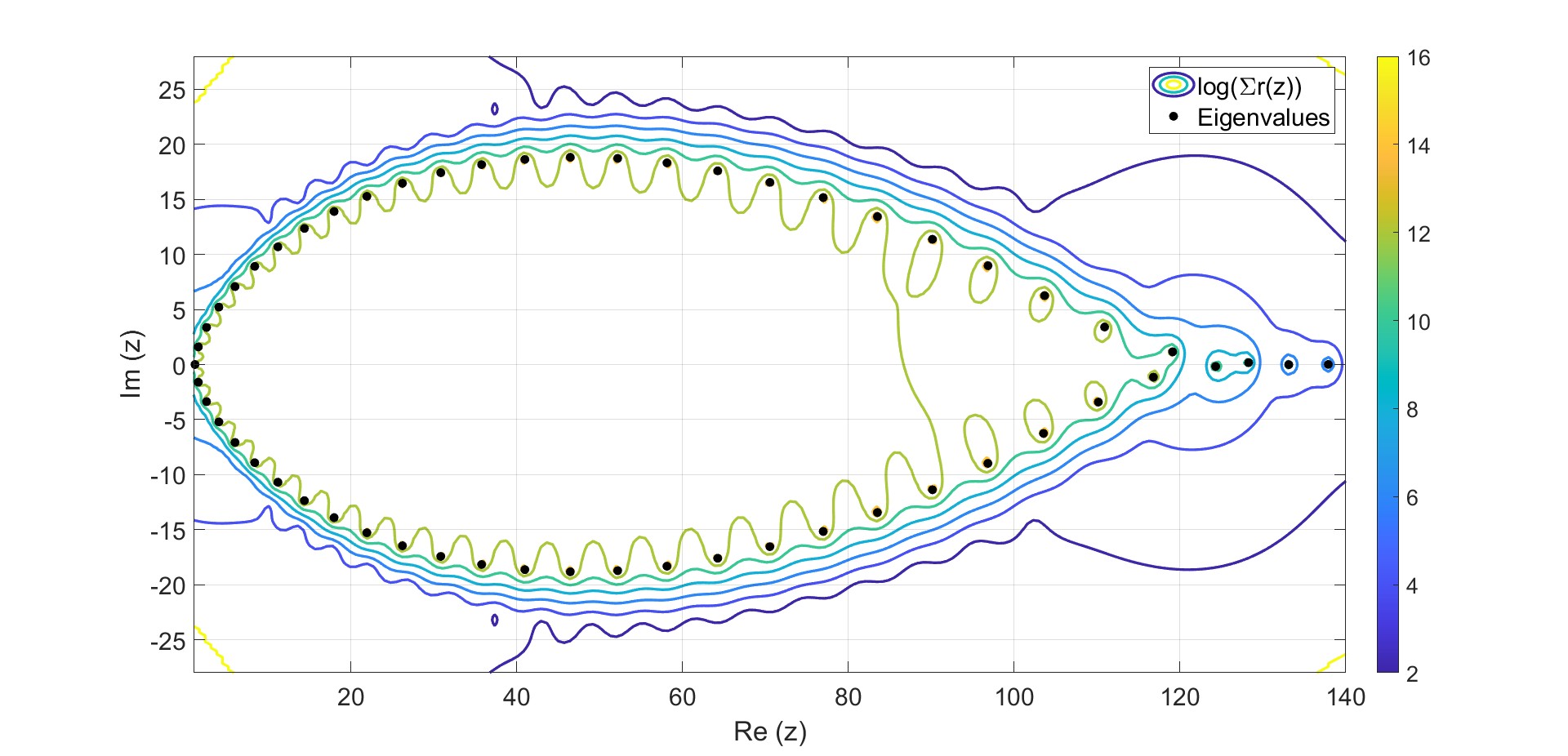} 
		\caption{ Relative variance $\Sigma_{r}$ in the complex plane for $N = 80$, $M = 107$, $\beta = 1$, $\alpha = 0.25$, and $\epsilon = 10^{-1}$, with averaging being performed over $10^{3}$ sample matrices. Black filled circles indicated positions of the roots of the mean characteristic polynomial. Top: Density plot of the logarithm  of $\Sigma_{r}$. Bottom: Contour plot of the logarithm of the relative variance.}
		\label{f:6}
	\end{center}
\end{figure}

 The upper panel of Figure \ref{f:6} shows a density plot of the logarithm of the relative variance $\Sigma_{r}$ in the complex plane, while the lower panel presents its corresponding contour plot. For clarity, the positions of the roots of the mean characteristic polynomial are also marked in both panels as black filled circles. We observe that the relative variance exhibits sharp peaks at the locations of these roots. Along the circumference, it drops off abruptly between neighboring roots, whereas in the direction perpendicular to the circumference, the decay is significantly more gradual. In the following subsection, we provide a quantitative analysis of this anisotropy.

	\subsection[Properties of relative variance]{Variance of the characteristic polynomial along two directions}

One observes in Figure \ref{f:5}, that  the equation $M_1(z)=0$ always has a root at $z = 0$.  To quantify the anisotropy in the behavior of the relative variance, we focus specifically on the region near this particular root and consider separately the forms of the relative variance along the real and imaginary axes. While a similar analysis could, in principle, be performed for all other roots, it becomes increasingly more involved because of the shape of the balloon-like structure on which the eigenvalues are located,  at the same time without offering much of an additional insight.
 For a large but finite value of $N$,  considering a small variation along the imaginary axis we can set $z=it$, where $t$ is a real small but bounded away from zero parameter, while setting the real part of $z$ equal to zero. Using the definitions from the previous subsection, linearizing the second moment  we get the following representation

\begin{equation}
		M_{2}(t)\approx \sum\limits_{i,j=N-1}^{N} \langle f_{N}^{i}f_{N}^{j} \rangle (-it)^{N-i}(it)^{N-j}=\langle  f_{N}^{N-1}f_{N}^{N-1} \rangle t^{2}+\langle f_{N}^{N}f_{N}^{N} \rangle, 
\end{equation}
 where the symbol $\approx$ indicates that only the leading terms in the summation are being considered.
Using next the definition of the first moment, we get

\begin{equation}
	M_{1}(t)=\prod\limits_{i=1}^{N}(it-\lambda_{i})=\prod\limits_{i=1}^{k}(it-\lambda_{i})\prod\limits_{j=k+1}^{N}(it-\lambda_{j}^{'}),
\end{equation}
where $\lambda^{'}$  represents complex eigenvalues.  Consequently, we find that the first order in $t$ the first moment obeys
\begin{equation}
	M_{1}\approx(-1)^{k-1}(it)\left[\prod\limits_{l=1}^{k} \lambda_{l}\right]\prod\limits_{j=1}^{\frac{N-k}{2}}\left[|\lambda_{j}^{'}|^{2}-it(\lambda_{j}^{'}+\overline{\lambda_{j}^{'}})\right],
\end{equation}
such that
\begin{equation}
	M_{1}\overline{M_{1}}\approx \left[\prod\limits_{i=2}^{N}\lambda_{i}\right]^{2}t^{2} .
\end{equation}
 From the above expressions we readily find that the variation of the relative variance along the imaginary axis follows
\begin{center}
	$\displaystyle \Sigma_{r}^{\mbox{im}}(t)$$\approx$ $\frac{\langle f_{N}^{N}f_{N}^{N} \rangle}{\left[\prod\limits_{i=1}^{N}\lambda_{i}\right]^{2}t^{2}}$,
\end{center}
with $\langle f_{N}^{N}f_{N}^{N} \rangle = \langle \left[\prod\limits_{i=1}^{N}\mu_{i}\right]^{2} \rangle$.\\
 For the variation along the real line, (with ${\rm Im(z)=0}$), we set $z=t$, where $t$ is a small real parameter, to get
\begin{equation}
	M_{1}(t) \approx \prod\limits_{i=1}^{k}(t-\lambda_{i})\prod_{j=1}^{\frac{N-k}{2}}\left( t^{2}-t(\lambda_{j}^{'}+\overline{\lambda^{'}_{j}})+|\lambda^{'}_{j}|^{2} \right) , 
\end{equation}
and
\begin{equation}
	M_{2}\approx \sum\limits_{i,j=N-1}^{N} \langle f_{N}^{i}f_{N}^{j} \rangle (-t)^{2N-i-j}\approx -2t\langle f_{N}^{N-1}f_{N}^{N} \rangle+\langle f_{N}^{N}f_{N}^{N} \rangle .
\end{equation}
   Noticing also that $M_{1}(t)=\overline{M_{1}(t)}$, we have
\begin{equation}
	M_{1}(t)\overline{M_{1}(t)}\approx \left[\prod\limits_{i=2}^{N}\lambda_{i}\right]^{2}t^{2}.
\end{equation}
Combining the above results, we find that the variation of  the relative variance along the real axis obeys, in the vicinity of the root $z = 0$,
\begin{center}
	$\displaystyle \Sigma_{r}^{\mbox{real}}(t)$$\approx$ $\frac{-2\langle f_{N}^{N-1}f_{N}^{N} \rangle}{\left[\prod\limits_{i=2}^{N}\lambda_{i}\right]^{2}t}$+$\frac{\langle f_{N}^{N}f_{N}^{N} \rangle}{\left[\prod\limits_{i=2}^{N}\lambda_{i}\right]^{2}t^{2}}$, 
\end{center}
 showing that the difference of the relative variances along the real and imaginary axes, in the vicinity of the root $z=0$, is of order 
\begin{center}
	$\displaystyle \Sigma_{r}^{\mbox{real}}-\Sigma_{r}^{\mbox{im}} \approx \frac{1}{|t|}$ \,.
\end{center}
Because $t$ is assumed to be small (but bounded away from zero), this difference can be therefore very large revealing a significant anisotropy in the behavior of fluctuations. This behavior is, of course, not unexpected. Fluctuations in the positions of eigenvalues along the circumference of the balloon-like structure (for the root $z=0$ it is $\Sigma_{r}^{\mbox{im}}$) are naturally suppressed due to the tendency of the eigenvalues to crystallize. In contrast, in the direction perpendicular to the circumference, (for the root $z=0$ it is $\Sigma_{r}^{\mbox{real}}$)
the eigenvalues have more freedom to fluctuate, leading to a significantly broader spread in their values.

\subsection{Self-Averaging via Rescaled Variables}	
\begin{figure}[hbt!]
		\begin{center}
			\includegraphics[width=0.7\textwidth]{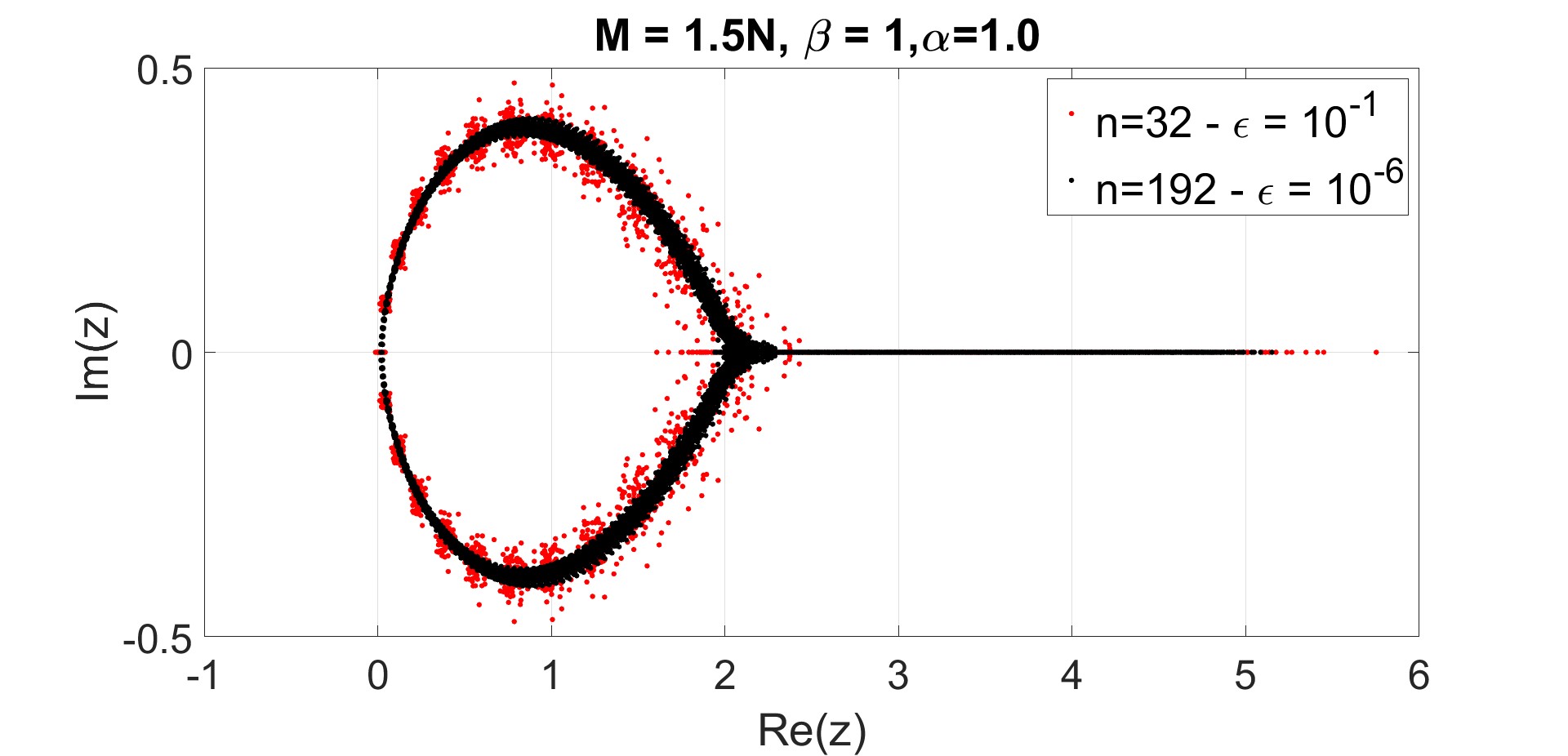}
			\caption{Eigenvalue distribution with parameters $M = 1.5 N$, $\beta = 1$, $N = 32$ and $\epsilon = 10^{-1}$: red dots; for  $N = 192$ and $\epsilon = 10^{-6}$: black dots. This allows evaluation of the form of the perturbation at the corners of the matrix.}
			\label{f:7} % <-- agora está dentro e depois do \caption
		\end{center}
	\end{figure}

	In this section, we present numerical results that explicitly demonstrate the self-averaging behavior. Let us begin with the distribution given in eq. (\ref{disF}) and perform the change of variable $y=x/\sqrt{N}$. This  introduces an explicit dependence on the matrix size $N$ into the joint probability distribution function, yielding
	\begin{equation}
		f_{\nu}(y)=\frac{2^{1-\frac{\nu}{2}}}{\Gamma(\nu/2)}y^{\nu-1}N^{\frac{\nu}{2}}\exp\left[-\frac{Ny^{2}}{2}\right], \quad x\geq 0
	\end{equation}
	This corresponds to multiplying the bidiagonal matrices $B_{\pm}$ by $\frac{1}{\sqrt{N}}$. Repeating the procedure described above, the construction of a tridiagonal matrix from bidiagonal ones, we obtain a new tridiagonal matrix $L'(\alpha,N)=\left[(1/\sqrt{N})B_{-}\right]\left[(1/\sqrt{N})B_{+}^{T}\right]$,
	
	\[
	L'(\alpha,N)=
	\left[\begin{array}{cccc}
		\frac{a_{N}^{2}}{N} & \frac{a_{N}b_{N-1}^{1+\alpha}}{\sqrt{N}^{2+\alpha}} & \dots & 0 \\
		\frac{a_{N}b_{N-1}^{1-\alpha}}{\sqrt{N}^{2-\alpha}}	
		& \frac{a_{N-1}^{2}+b_{N-1}^{2}}{N} & \ddots & \vdots \\
		\vdots & \ddots & \ddots & \frac{a_{2}b_{1}^{1+\alpha}}{\sqrt{N}^{2+\alpha}}	
		\\
		0 & \dots & \frac{a_{2}b_{1}^{1-\alpha}}{\sqrt{N}^{2-\alpha}}	
		
		& \frac{a_{1}^{2}+b_{1}^{2}}{N}
	\end{array}\right],
	\]
	which depends explicitly on $N$. Within such a  formulation, we turn to the large-$N$ limit. As in the original framework, we perturb the matrix $L’$ by $\epsilon$ at the corners. Since the diagonal matrix $\eta$ is now given by,
	\begin{equation}
		\label{eta}
		\mbox{diag}(\eta)=\left(1,\frac{b_{N-1}^{2\alpha}}{\sqrt{N}^{2\alpha}},\dots,\prod\limits_{k=1}^{N-1} \frac{(b_{N-k})^{2\alpha}}{\sqrt{N}^{2\alpha}}\right),
	\end{equation}
	using the definition of the matrix $\eta^{1/2}$ given above, this leads to our final matrix $H’_{2}$, which will be used to compute the eigenvalues numerically. Explicitly, 
	\[
	H'_{2}(\alpha,\epsilon,N)=\frac{1}{N}
	\left[
	\begin{array}{cccc}
		a_{N}^{2}  & a_{N}b_{N-1}  &  & Q'_{2}  \\
		a_{N}b_{N-1}  & a_{N-1}^{2}+b_{N-1}^{2} & \ddots  &  \\
		& \ddots  & \ddots  & a_{2}b_{1} \\
		Q'_{1} &  & a_{2}b_{1} & a_{1}^{2}+b_{1}^{2}
	\end{array}
	\right],
	\]
	where $Q'_{1}=\epsilon N^{\frac{-\alpha}{2}(N-1)} \prod\limits_{i=1}^{N-1} (b_{N-i})^{\alpha}$ and $Q'_{2}=\epsilon N^{\frac{\alpha}{2}(N-1)} \prod\limits_{i=1}^{N-1} (b_{N-i})^{- \alpha}$.\\
	Before expanding the determinant, it is necessary to evaluate the form of the perturbations at the corners of the matrix. Based on numerical results, Figure (\ref{f:7}) presents the eigenvalues for different values of $N$ and $\epsilon$. The red dots correspond to the case $N = 32$ and $\epsilon=10^{-1}$, while the black dots represent the eigenvalues of matrices with parameters $N = 192$ and $\epsilon=10^{-6}$. From these results, we observe that the perturbation can be assumed to take the form
	\begin{equation}
		\epsilon_{N}=\epsilon_{0}^{N},
	\end{equation}
	where for the matrices in Figure (\ref{f:7}), we have $\epsilon_{0}^{N/32} $. 
	
	With this setup, we compute the characteristic polynomial of $H’_{2}$. To achieve this, we note that the term to be neglected is the one in the left corner. Expanding the determinant as it was done above, we can write
	\begin{equation}
		\label{newPN}
		\langle P_{N}(N\lambda / \beta)\rangle=\langle N^{1+(N-1)\alpha/2}\epsilon\prod_{i=0}^{N-2} a_{N-i} \prod_{j=1}^{N-1} b_{N-j}^{1-\alpha} \rangle,
	\end{equation}
	where $\langle P_{N}(N\lambda / \beta)\rangle=N!\beta^{N}L_{N}^{M-N}(N\lambda / \beta)$, $\langle  a_{N-i}\rangle = [(M-i)\beta/2]^{1/2}$ and $\langle  b_{N-j}^{1-\alpha}\rangle = [(N-i)\beta/2]^{(1-\alpha)/2}$. This procedure formally demonstrates that the framework developed earlier remains valid, provided that the appropriate modifications are applied. Comparing Eq. (\ref{newPN}) with the previous expression, Eq. (\ref{laguerre}), reveals that the essential difference lies in the argument of the polynomial as well as in the perturbative term, as can also be inferred from Eq. (\ref{originalQ}). These results complete the discussion initiated in the previous section and offer additional evidence in support of the proposed self-averaging behavior.
	\begin{figure}[h]
		\centering
		\includegraphics[width=0.9\linewidth]{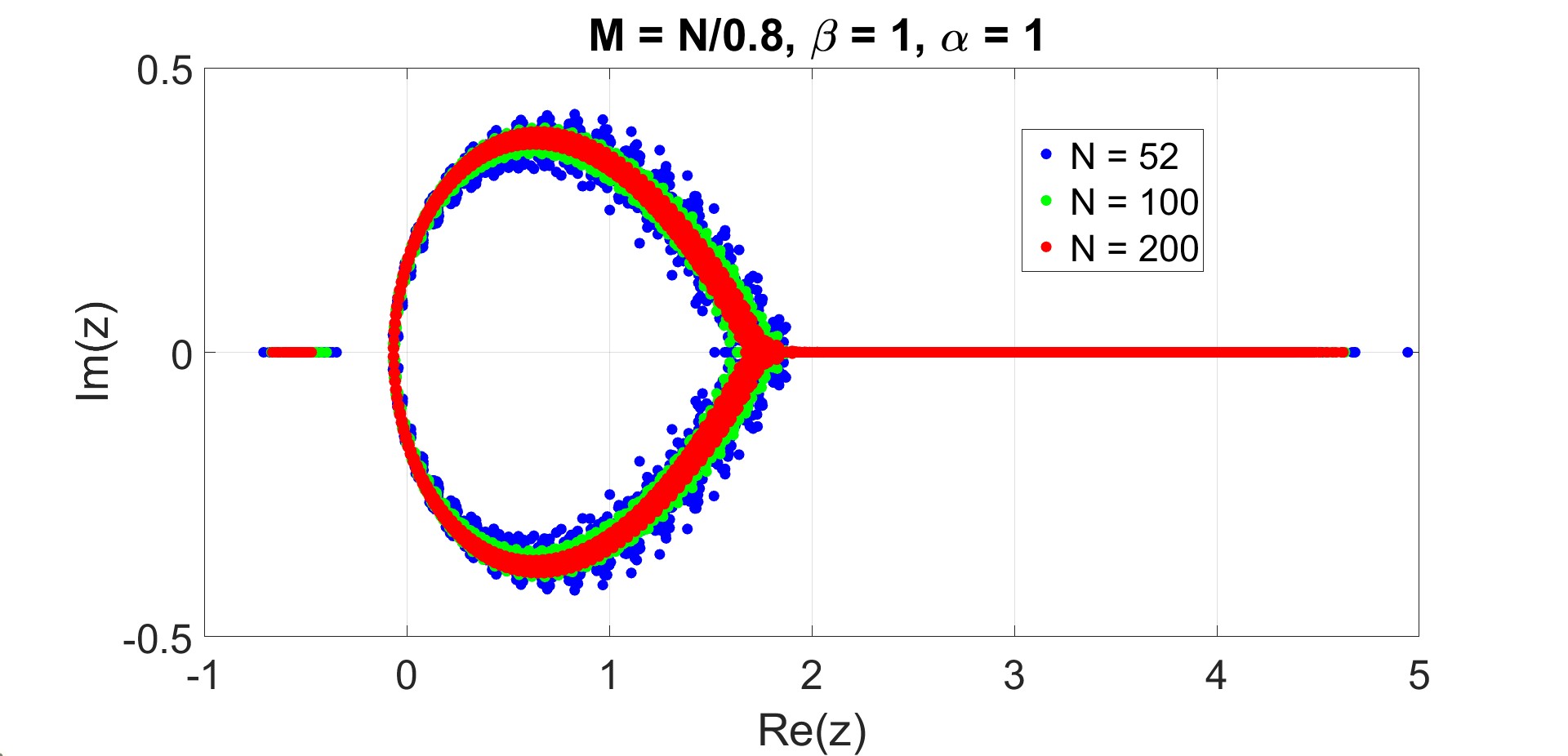}
		\caption{Distribution of eigenvalues in the complex plane for rescaled matrices $H’_{2}(\alpha,\epsilon, N)$ with parameters $M=N/0.8$, $\beta = 1$, $\epsilon = 1$, and $100$ samples. Results are shown for three different matrix sizes: $N=52$ (blue), $N=100$ (green), and $N=200$ (red).}
		\label{rescale}
	\end{figure}
	 Figure (\ref{rescale}) presents a representative numerical result. The distribution of eigenvalues in the complex plane is shown for $M=N/0.8$, $\beta = 1$, $\epsilon = 1$, and three matrix sizes, $N$, $N=52, 100, 200$. The characteristic “balloon” structure discussed throughout the paper is clearly visible, together with the corresponding fluctuations. A key observation is that these fluctuations decrease as $N$ increases. In the large-$N$ limit, the results indicate that the fluctuations are suppressed, suggesting that the spectrum approaches convergence.

\section{Conclusion}

 Non-Hermiticity has emerged as a powerful analytical framework for exploring a wide range of phenomena across various physical systems, particularly in the context of open quantum systems. Beyond its physical relevance, the intricate structure of the eigenvalues associated with non-Hermitian operators is of considerable interest in its own right. 

In this paper, we introduced a pseudo-Hermitian model constructed from $\beta$-Laguerre random matrices, incorporating an unbounded diagonal auxiliary matrix $\eta$ (see eq.  (\ref{eta})). The resulting ensemble has a highly unstable spectrum: even a small perturbation breaks the system's symmetry, causing the eigenvalues to shift from the real axis into the complex plane. Our primary goal was to analytically determine the loci of these eigenvalues in the complex plane within such a non-Hermitian ensemble.

While such analyses are often carried out by averaging the resolvent, this approach appears analytically intractable for the present model. However, it is well known that in many settings—particularly for Hermitian matrices— the resolvent becomes deterministic in the large-dimension limit due to self-averaging behavior. Motivated by this, we first presented numerical evidence suggesting that a similar self-averaging phenomenon arises in the perturbed $\beta$-Laguerre ensemble defined in eq. (\ref{main}). Building on this observation, we analytically determined the curve in the complex plane along which the eigenvalues lie by examining the roots of the mean characteristic polynomial. Our results reveal that this curve forms the boundary of a closed, anisotropic, balloon-like structure: smooth at the left extremity and featuring a cusp singularity at the right. This cusp is followed by a finite discrete set of real eigenvalues. Numerical simulations confirmed excellent agreement with our theoretical predictions, even for relatively modest matrix sizes. Furthermore, we showed that in the large-dimension limit, this curve approach to a circle—an observation again supported by numerical evidence. 

To further validate the self-averaging hypothesis, we analyzed the second moment of the characteristic polynomial and evaluated its relative variance. Using the rescaled formulation, last section, we demonstrated that, in the large-matrix limit, fluctuations are strongly suppressed, leading to a clear concentration around the mean. Moreover, we identified a pronounced anisotropy in the fluctuation behavior: while fluctuations are strongly reduced along the circumference of the balloon-like structure, they become markedly more pronounced in the perpendicular direction. These results provide compelling evidence that the system exhibits self-averaging.

The present work opens several avenues for future research.  One interesting direction is to extend the saddle-point approximation to study the influence of Gaussian distributions in all classical $\beta$-ensembles made pseudo-Hermitian.
In the context of complex eigenvalues, the effect of sign changes between the two off-diagonal elements is particularly noteworthy, as it leads to a two-dimensional spectral distribution—a topic of active interest in the literature. Furthermore, extending this study to the classical $\beta$-Jacobi (MANOVA) ensemble would be valuable, contributing to a more comprehensive understanding of all classical ensembles in random matrix theory.

\section*{Acknowledgments}

C. A. Goulart  acknowledges a financial support from the Brazilian agency CAPES under the grant number 88887.937023/024-00 and also wishes to thank the Laboratoire de Physique Th\'eorique de  la Mati\`{e}re Condens\'ee, Sorbonne Universit\'e, for a warm hospitality during his six months long internship in $2024$, when this work has been initiated. The authors wish to thank L. A. Pastur for helpful discussions.

\newpage
\section*{References}
\bibliographystyle{ieeetr}
\bibliography{ref-2}	

\end{document}